\documentstyle[prb,preprint,aps]{revtex}
\begin{document}
\title {Microscopic analytical theory of a correlated, two-dimensional
\\ $N$-electron gas in a magnetic field} %
\author{Neil F. Johnson$^{1,*}$ and Luis
Quiroga$^2$} %
\address {$^1$Department of
Physics, Oxford University, Parks Road, Oxford, OX1 3PU,  England} \address
{$^2$Departamento de Fisica, Universidad de Los Andes, Bogota,  Apartado Aereo
4976,  Colombia}
\address {$^*$ E-mail: n.johnson@physics.oxford.ac.uk}
\date{\today}
\maketitle

\begin{abstract} We present a microscopic, analytical theory describing a confined
$N$-electron gas in two dimensions subject to an external magnetic field.  The
number of electrons $N$ and strength of the electron-electron interaction can be
arbitrarily large, and all Landau levels are included implicitly.
For any value of the magnetic field
$B$, the correlated
$N$-electron states are determined by the solution to a universal effective
problem which resembles a ficticious particle moving in a multidimensional space,
without a magnetic field, occupied by potential minima
corresponding to the classical $N$-electron equilibrium configurations.
Introducing the requirement of total wavefunction antisymmetry selects out the
allowed minimum energy $N$-electron states. 
It is shown that low-energy minima
can exist at filling factors $\nu=\frac{p}{2n+1}$ where $p$ and $n$ are any 
positive
integers. These filling factors correspond to the experimentally observed
Fractional (FQHE) and Integer  (IQHE) Quantum Hall effects. The energy gaps
calculated analytically at $\nu = \frac{p}{3}$ are found to be consistent with
experimental data as a function of magnetic field, over a range of
samples.


\end{abstract} 

{\bf I. INTRODUCTION}

The problem of a highly-correlated, two-dimensional electron gas in an external
magnetic field has attracted much attention in the past decade. Of particular
interest is the microscopic origin of the observed fractions in the Fractional
Quantum Hall Effect (FQHE) \cite{FQHE,alan,Chakraborty}. In the past few years, it
has also been appreciated that many-body effects play a role in the formation of the
gaps giving rise to the Integer Quantum Hall Effect (IQHE). As a complement to the
experimental work in this subject, there have been many theoretical models
proposed for both the FQHE and the IQHE. These range from field-theoretical
treatments through to numerical, finite-size ($N\leq 6$) calculations. One of the
most successful theoretical developments has been the proposal of trial
wavefunctions by Laughlin and others \cite{FQHE,Chakraborty,Laughlin,bert} to
describe the interplay of wavefunction antisymmetry and electron-electron repulsion
that effectively allows electrons in the lowest Landau level to form a highly
correlated electron liquid. A related development by Jain \cite{jain} considers  the
construction of `composite' fermions by attaching flux tubes to each electron --
recent work on Chern-Simons field theories provides some support for such
composite-fermion construction schemes
\cite{Chakraborty,HLR}.
The general problem of describing an
$N$-electron gas in an external magnetic field has recently taken on additional
importance in  semiconductor  physics  due to the fabrication of quantum dots
containing  a finite number of electrons  \cite{ashoori,mceuen,Maksym}. It
is interesting to note that although the FQHE was originally observed in infinite
two-dimensional electron gases (2DEG), it even persists in quantum dots containing
a large but finite number $N$ of  electrons \cite{hansen}.   

Given the fact that the underlying, microscopic $N$-electron Hamiltonian is known,
one could ask whether there exists an alternative, more direct way of
understanding the nature of highly-correlated electron states  {\em without}
recourse to composite-fermion constructions, effective field theories,
restrictions to lowest Landau levels and/or small numbers of electrons. The obvious
stumbling blocks are that the electron-electron repulsion and the cyclotron energy
are typically comparable in magnitude, and that $N$-electron Schrodinger equations
are generally intractable analytically. 

In this paper we pursue such an alternative approach starting with an
$N$-electron Schrodinger equation. We develop a microscopic,
analytical theory describing correlated states of a confined
$N$-electron gas in two dimensions subject to an external magnetic field $B$.  The
number of electrons $N$ and the strength of the electron-electron interaction can
be arbitrarily large, and all Landau levels are included implicitly. We show that
the description of $N$-electron correlated states at finite
$B$ reduces to a universal effective problem which resembles a ficticious particle
moving in a multidimensional space occupied by potential minima corresponding to
the classical $N$-electron equilibrium configurations. Introducing the requirement
of $N$-electron wavefunction antisymmetry selects out the allowed minimum energy
$N$-electron states. 
A possible connection with the FQHE and IQHE is then proposed. In particular, it is 
argued
that low-energy minima can form at particular
angular momenta corresponding to filling factors $\nu=\frac{p}{2n+1}$ where $p$
and $n$ are any integer. These filling factors correspond to those observed
experimentally in the FQHE and IQHE.  

The present theory suggests the following possible physical
interpretation of FQHE and IQHE states. Consider 
an $N$-electron wavefunction localized around a Wigner crystal
(WX) configuration with total relative angular momentum $J$. At particular
values of $J$, $N$-electron wavefunctions localized around nearby
defect configurations (i.e. WX plus defect which we shall denote as WXD) can
co-exist; we note that the allowed values of $J$ such that $N$-electron states
can co-exist around WX and WXD simultaneously are severely restricted by the
requirement of total wavefunction antisymmetry. At these common $J$ values,
which we shall denote as $J=J_m$, hybridization of the $N$-electron
states centered on the WX and WXD minima can occur. This hybridization
effectively allows the electrons in the WX solid to diffuse  throughout the system
via WXD defect states. The resulting delocalized  `liquid'-like $N$-electron
state has a lower zero-point energy -- a gap therefore opens up between
the  liquid-like states at $J=J_m$ and other states at $J\neq J_m$. For large
$N$ the resulting liquid-like ground-states at $J=J_m$  have filling 
factors given by
the well-known formula \cite{Chakraborty}  $\nu=\frac{N(N-1)}{2J_m}$. {\em We find
that the $\nu$ values at which these gaps arise are  identical to those observed
experimentally in the  FQHE and IQHE}.  The energy gaps calculated analytically
at $\nu = \frac{p}{3}$ are found to
 be consistent with experimental data
obtained from a range of samples. Various other known features of FQHE states
can  also be
reproduced.  

The model avoids discussion of one-electron properties such
as Landau levels,  and therefore offers the possibility of a unified description
of both the FQHE and IQHE based on a  microscopic $N$-electron Schrodinger
equation. The formalism in this paper builds on an earlier model presented by us 
in Ref. 12. In particular, we conjectured in Ref. 12 that the
classical minimum energy configurations play a crucial role in deciding the
symmetry-allowed $N$-electron correlated states in few-electron quantum dots. 
It was pointed out that the
classical minimum energy configurations for $N<6$ all  consist of  $N$ particles
on a  ring, while for $N=6$ additional minima occur \cite{bolton,Bedanov}.
Curiously,  it is precisely at $N=6$ that the magic number $J$  sequence of $\Delta
J=N$ is  broken.  This idea was independently pursued by Maksym in a fascinating
way \cite{Eckardt} for
$N\leq 6$ -- the classical Eckardt frame was employed to study correlated
few-electron dynamics and, in particular, the possible existence of `liquid'-like
states. We note that the term `liquid'-like was introduced by Maksym to describe
the loss of symmetry occuring when states corresponding to different classical
minima are allowed to mix. This terminology will also be used in the present
paper.  We wish to emphasize that the model presented here is qualitatively
different from an earlier theoretical approach of Kivelson et al.
\cite{Kivelson} based on the so-called cooperative ring exchange. In short,
we are suggesting here that FQHE states are the liquid-like states resulting from the
hybridization of $N$-electron wavefunctions localized around both crystal (WX)
and  crystal-plus-defect configurations (WXD). 

The outline of the paper is as follows. In Sec. II we present the microscopic
$N$-electron Schrodinger equation. The hyperangular coordinate system is 
introduced for the relative motion Hamiltonian. The problem then reduces to a
$2N-4$ hyperangular equation (Sec. II.1). In Sec. II.2 the specific case of $N=3$
is outlined. This was discussed in detail in Ref. 12 and is reviewed
here since it is useful for visualizing the $N$-electron results. In
Sec. II.3 a simplified hyperangular equation is obtained  which is valid in
the regime of strong electron-electron interactions and for large $N$. The
characteristics of the lowest energy solutions are discussed. Section III
addresses the requirement of $N$-electron wavefunction antisymmetry. Permutation
symmetries of the $N$-electron wavefunction become space-group operations in the
multi-dimensional hyperangular configuration space. The states which will become
ground states separated by a finite energy gap are found to correspond to filling
fractions observed in the FQHE. Section IV obtains analytic estimates for
the FQHE gaps at the fractions $\frac{p}{3}$ as an example. These
estimates are found to be consistent with experimental data over a range
of samples, despite the fact that the results emerge from a simple
one-dimensional, particle-in-a-box equation. Section V summarizes the results.    

\vskip 0.2in

{\bf II. MICROSCOPIC $N$-ELECTRON HAMILTONIAN}

The analytic tractability of our model is made
possible by a combination of a parabolic confinement potential and an inverse-square
electron-electron repulsion potential.  The parabolic confinement is known to be a
reasonable approximation for many semiconductor quantum dot samples \cite{review}. 
For the
case of a heterostructure  (i.e. 2DEG) it 
mimics the effect of a positive
background yielding an approximately uniform electron density in the large $N$
limit (see Ref. 14). 
The $\beta/r^2$ electron-electron interaction ($\beta>0$)  is not
unrealistic in quantum
dots due to the presence of image charges in neighboring electrodes. 
Recent theoretical work suggests \cite{Maksym96} that 
the true repulsive interaction between  electrons in
a quantum dot dot is more likely to be proportional to $1/r^n$ with $n\sim 3$ at
large $r$ and $n\sim 1$ at small $r$. 
In
heterostructures, the  electron-electron interaction is probably less affected by
image-charge effects. However the general features of our results, which are based on the
assumption that $n=2$ for all $r$, should still be qualitatively useful. 
In particular, the occurrence of FQHE in two-dimensional electron gases is not
thought to depend crucially on the precise form of the electron-electron
repulsion.  Recent  quantitative comparisons 
\cite{review,madhav,kinaret}  have indeed shown that the $1/r^2$ and $1/r$
repulsive interactions  yield $N$-electron energy spectra with very similar 
features.  Of particular relevance to the present theory is
the finding that the {\em classical} minimum energy configurations for $N$
electrons in a two-dimensional parabolic confinement potential seem to be almost
identical for both $1/r$ and $1/r^2$ interactions \cite{pakming}. 

The exact Schrodinger equation for $N$ electrons with repulsive  interaction
$\beta/r^2$ moving in a 2D parabolic potential  subject to a magnetic field $B$
(symmetric gauge) along the z-axis,  is given by $(H_{\rm space}+H_{\rm
spin})\Psi=E\Psi$;  
$$H_{\rm space}=\sum_{i=1}^{N} (\frac{({\bf
p}_i-\frac{e{\bf A}_i}{c})^2}{2m^*}  + {\frac{1}{2}}m^*\omega_0^2 |{\bf
r}_i|^2  )+  \sum_{i<j} {\frac{\beta}{|{\bf r}_i-{\bf r}_j|^2}}$$ 
\begin{equation} =\sum_{i=1}^{N} (\frac{{\bf
p}_i^2}{2m^*}  + {\frac{1}{2}}m^*\omega_0^2(B) |{\bf r}_i|^2  + \frac{\omega_c}{2}
l_i)+  \sum_{i<j} {\frac{\beta}{|{\bf r}_i-{\bf r}_j|^2}} \end{equation}  where
$\omega_0^2(B)=\omega_0^2+\frac{\omega_c^2}{4}$,  $\omega_c$ is  the cyclotron
frequency, and  $H_{\rm spin}=-{g^*\mu_B B}\sum_i s_{i,z}$.  The momentum and
position of the $i$'th particle are  given by 2D  vectors ${\bf p}_i$ and ${\bf
r}_i$ respectively; $l_i$  is the  z-component  of the  angular momentum.  The
exact eigenstates  are written in terms of products of spatial and spin 
eigenstates  obtained  from $H_{space}$ and $H_{spin}$ respectively. Eigenstates
of  $H_{\rm spin}$ are just products of the spinors of the  individual 
electrons and have energy $E_{spin}=g^*\mu_B B S_z$, where $S_z$ is
the 
$z$-component of total spin and $g^*$ is the electron effective $g$-factor.   We
employ standard Jacobi coordinates
${\bf X}_j$   ($j=1,2,\dots
,N$)  where   ${\bf X}_1=\frac{1}{N}\sum_j {\bf r}_j$ (center-of-mass coordinate) 
and ${\bf X}_{j>1}$ (relative coordinates) is given by
\begin{equation} 
{\bf
X}_j=\big[{\frac{j-1}{j}}\big]^{\frac{1}{2}}\big[{\bf r}_j-
\frac{1}{j-1}({\bf r}_1+ {\bf r}_2+\dots
{\bf r}_{j-1})
\big] \end{equation}   
together  with the conjugate momenta ${\bf P}_j$ (see
Fig. 1 for $N=3$).  The
center-of-mass motion decouples,  $H_{\rm space}=H_{\rm CM}({\bf X}_1)+ H_{\rm
rel}(\{{\bf X}_{j>1}\})$, hence $E_{\rm space}=E_{\rm CM}+E_{\rm rel}$. The exact
eigenstates of $H_{\rm CM}$ and energies $E_{\rm CM}$  are well-known 
\cite{fock}.  The non-trivial problem  is to solve the relative motion equation 
$H_{\rm rel}\psi=E_{\rm rel}\psi$. We transform the relative coordinates  $\{{\bf
X}_{j>1}\}$ to standard hyperspherical  coordinates: ${\bf
X}_j=r(\prod_{i=j+1}^{N}{\rm cos}\alpha_{i}){\rm sin} \alpha_{j}  e^{i\theta_j}$
with $r\geq 0$ and $0\leq\alpha_j\leq\frac{\pi}{2}$  ($\alpha_{2}=\frac{\pi}{2}$). 
Physically, the hyperradius $r$ is just the root-mean-square electron-electron
separation. The exact eigenstates of $H_{\rm rel}$ have the form $\psi_{\rm
rel}=R(r)F(\Omega)$ where $\Omega$ denotes the $(2N-3)$ hyperangular 
$\{\theta;\alpha\}$ variables; $R(r)$ and $F(\Omega)$ are  solutions of the
hyperradial and hyperangular  equations respectively.  The hyperradial equation is
given by
\begin{equation} \big[\frac{d^2}{d r^2}+ \frac{2N-3}{r}\frac{d}{d
r}-\frac{\gamma(\gamma+2N-4)}{r^2}- \frac{r^2}{l_0^4}+\frac{2m^*(E_{\rm rel}-
\hbar J\frac{\omega_c}{2})}{\hbar^2} \big] R(r)=0 \end{equation} where
$l_0^2=\hbar(m^*\omega_0(B))^{-1}$ and $J$ is the total {\em relative} 
angular momentum. The parameter  $\gamma>0$ and is related to
the  eigenvalue of the $B$ and $\omega_0$-independent hyperangular equation 
(see Sec. II.1).  Equation (3) can be solved exactly yielding \begin{equation}
E_{\rm rel}=\hbar\omega_0(B)[2n+\gamma+N-1]+ J\frac{\hbar\omega_c}{2} \end{equation}
where $n$ is any positive integer or zero and  \begin{equation}
R(r)=\big[\frac{r}{l_0}\big]^\gamma L_n^{\gamma+N-2} \big(\frac{r^2}{l_0^2}\big)
e^{-\frac{r^2}{2l_0^2}}\ \ . \end{equation}  Equation (4) provides an exact and
infinite set of relative breathing-mode excitations $2\hbar\omega_0(B)\Delta n$  
for any 
$N$ regardless of particle statistics and/or  spin states. These quantum breathing modes
were first reported in Ref. 12, and later confirmed by Geller et al. \cite{Geller} -- the
classical version of these modes for the Coulomb interaction was discussed in detail by
Schweigert et al. \cite{Schweigert}. 

\vskip 0.2in 
{\bf II.1 Exact hyperangular equation for any $N$}

It remains to solve  the   $B$ and $\omega_0$-independent hyperangular 
equation which is given by
\begin{equation} \big[\Theta^2_N+\frac{2m^*\beta}{\hbar^2}V(\Omega)\big]
F(\Omega) =[\gamma(\gamma+2N-4)]F(\Omega) \end{equation} where
\begin{equation} \Theta^2_N\equiv -\frac{\partial^2}{\partial \alpha_N^2}
+\frac{[2N-6-(2N-4){\rm cos}2\alpha_N]}{{\rm
sin}2\alpha_N}\frac{\partial}{\partial \alpha_N}+{\rm
sec}^2\alpha_N\Theta^2_{N-1}-{\rm
cosec}^2\alpha_N\frac{\partial^2}{\partial\theta_N^2} \ . \end{equation} The
quantity $V(\Omega)$ is given by $r^2\sum_{i<j}|{\bf r}_i-{\bf r}_j|^{-2}$
and only depends on hyperangular coordinates $\Omega\equiv\{\theta;\alpha\}$. We
emphasize that this hyperangular equation (Eq. (6)) is universal in that it is
independent of the values of the magnetic field or confinement: solving Eq. (6) for
$\gamma$, and hence using Eq. (4), yields the complete solutions of the
$N$-electron Hamiltonian $H$ for all magnetic fields and confinement strengths.
Unfortunately the hyperangular equation does not admit exact
solutions for $\gamma$.  Sections II.2--4 will consider various approximations to
Eq. (6) which make the problem tractable.  Because $J$ remains a good quantum
number, we  can introduce a Jacobi transformation of the relative motion angles
$\{\theta_i\}$:  in particular  
\begin{equation}
\theta'=\frac{1}{N-1}\sum_{j=2}^N \theta_j \end{equation} and \begin{equation}
{\theta}_{[j]}=\big[{\frac{j-2}{j-1}}\big]^{\frac{1}{2}}\big[{\theta}_j-
\frac{1}{j-2}({\theta}_2+ {\theta}_3+\dots {\theta}_{j-1})
\big]
\end{equation}    where $j=3,4\dots N$. We hence have one $\theta'$
variable, $(N-2)$  $\theta_{[j]}$ variables,  $(N-2)$ $\alpha$ variables and one
hyperradius $r$ giving a  total of $2N-2$ variables as required for the relative
motion.  The exact eigenstates of $H_{\rm rel}$ have the form
$\psi=e^{iJ\theta'}R(r)G(\Omega')$ where $\Omega'$ denotes  the $(2N-4)$ 
$\{\theta_{[j]};\alpha_j\}$ variables excluding $\theta'$. The term $V(\Omega)$
is independent of $\theta'$ and will hence be written as $V(\Omega')$. It is
useful  to rewrite the eigenvalue of the hyperangular equation in terms of a new
variable $\epsilon$ as follows: \begin{equation}
\epsilon=\frac{\hbar^2}{8}\big[\gamma(\gamma+2N-4)-J^2- 
\frac{2m^*\beta}{\hbar^2}V(\Omega'_0)\big] \end{equation} 
where $V(\Omega'_0)$ is the value of $V(\Omega')$ evaluated at the hyperangles
corresponding to a particular classical, minimum-energy  $N$-electron
configuration (Wigner molecule). Permuting electron indices will
provide a set ${\{\Omega'_i\}}$ of {\em symmetrically equivalent minima} (SEM)
\cite{term,Eckardt} with the same potential energy 
$V(\Omega'_i)\equiv V(\Omega'_0)$ for all $i$ (e.g. ${\Omega'_0}$ and
${\Omega'_1}$ shown schematically in Fig. 2). Such SEMs have the same
topological structure but cannot be transformed into each other by 
rotations
\cite{term,Eckardt}. As will be shown in Sec. II.2, there are two such SEMs for
$N=3$.  The quantity
$\epsilon$ in Eq. (10) accounts for the contribution to the eigenvalue of the
hyperangular equation {\em without} including either the contributions from the
rigid-body rotational energy
$J^2$ or the electrostatic potential energy 
$\frac{2m^*\beta}{\hbar^2}V(\Omega'_0)$ of the classical minimum-energy
configuration.
 Physically therefore, $\epsilon$
contains the zero point energy in $\Omega'$-space   associated with the
quantum-mechanical spread of  $G(\Omega')$  about the minima $\{\Omega'_i\}$. The
actual spread in $G(\Omega')$ and hence $\epsilon$  will depend on the total
wavefunction antisymmetry  requirement. This is illustrated for $N=3$ in Sec. II.2
and discussed for large $N$ in Secs. II.3, II.4 and III. In
general $\epsilon\geq 0$,   $\epsilon\sim \beta^\mu$ where $\mu<1$, and
$\epsilon\sim J^\delta$  where $\delta<2$: these statements will be illustrated
in Sec. II.2 for $N=3$.  It is
straightforward to show that the term $\frac{2m^*\beta}{\hbar^2}V(\Omega'_0)$
appearing in the definition of $\epsilon$ is identical to $[\frac{V_{\rm
class}}{\hbar\omega_0(B)}]^2$ where $V_{\rm class}$ is the potential energy of 
the classical, minimum-energy $N$-electron configuration, thereby recovering the
expression given in Ref. 12. Note that 
  $V_{\rm class}\propto\beta^{\frac{1}{2}}\omega_0(B)$ and  that 
$\epsilon$ (like $\gamma$) is independent of $B$ and  $\omega_0$.  
The exact relative energy for any $N$ can now be written as 
\begin{equation} E_{\rm rel}=\hbar\omega_0(B)\big[2n+
\big([N-2]^2+J^2+\frac{2m^*\beta}{\hbar^2}V(\Omega'_0)
+\frac{8\epsilon} {\hbar^2}\big)^{\frac{1}{2}} +1\big]\\ 
+J\frac{\hbar\omega_c}{2}\ \ .  \end{equation} 
$E_{\rm rel}$ only depends on particle
statistics  through $\epsilon$.  As $\hbar\rightarrow 0$, $\epsilon\rightarrow 0$
and $E_{\rm rel}\rightarrow V_{\rm class}$.  

The exact $E_{\rm rel}$ expression has an important
consequence. The $J$-dependence of $\epsilon$ is weaker than $J^2$  as
$J\rightarrow 0$. Hence the term $J\frac{\hbar\omega_c}{2}$ in $E_{\rm rel}$ will
dominate the $J$-dependence of $E_{\rm rel}$ for small $J$ at fixed magnetic
field $\omega_c$. For states with $J<0$, $E_{\rm rel}$ will hence initially
decrease as $|J|$ increases. On the other hand at large negative $J$, $E_{\rm
rel}$ will tend to  $\hbar(\omega_0(B)-\frac{\omega_c}{2})|J|$ and hence will
increase linearly with $|J|$ at a given $\omega_c$. This implies that $E_{\rm
rel}$ has a {\em minimum} at a finite negative $J$ for a given fixed magnetic
field $\omega_c$. This is the basic mechanism behind the tendency of an
$N$-electron gas to form ground states at increasingly large $J$ values as the
magnetic field is increased. As will be shown in Sec. II.2 for $N=3$ electrons,
and in Sec. III for large $N$, only a subset of these $J$ minima are permitted
under the requirement of total wavefunction antisymmetry. These $J$-values are
often called `magic number' $J$-values in the context of few-electron quantum
dots. In Sec. III we will show that the analogous `magic number' $J$ states for a
large-$N$ electron gas  constitute FQHE and IQHE states. We emphasize that so far
our results are exact for any electron number $N$, electron-electron interaction
strength $\beta$, magnetic field $\omega_c$ and parabolic confinement $\omega_0$. 

\vskip 0.2in 
{\bf II.2 Specific case of $N=3$}

This case was studied in Ref. 12. Here we will summarize the
results since they are important for understanding the general $N$ case. For
convenience we change variables from $\alpha,\theta$ (c.f. Fig. 1)  to  $x,y$ where 
$x={\rm ln}[{\rm tan}(\frac{\pi}{2}-\alpha)]$ and  $y=\theta-\frac{\pi}{2}$. Since
$0\leq\alpha\leq\frac{\pi}{2}$, hence  $-\infty\leq x\leq\infty$ (N.B. $-\pi\leq
y\leq\pi$).  We define  $p_x=\frac{\hbar}{i}\frac{\partial}{\partial x}$ and
$p_y=\frac{\hbar}{i}\frac{\partial}{\partial y}$.  The exact hyperrangular
equation (Eq. (6)) now takes the  form \begin{equation}
\big[\frac{p_x^2}{2}+\frac{(p_y+\frac{\hbar  J {\rm cos}(2{\rm
tan}^{-1}e^x)}{2})^2} {2}\\ + V(x,y;\epsilon)\big]G(x,y)=\epsilon G(x,y)
\end{equation}   where  \begin{eqnarray}
 V(x,y;\epsilon)=m^*\beta\big[ \frac{(2+{\rm cos}(2{\rm tan}^{-1}e^x)}{({\rm cosec}
(2{\rm tan}^{-1}e^x)+ {\rm cot}({\rm tan}^{-1}e^x))^2- 3{\rm sin}^2 y} -
\frac{3}{4}{\rm sin}^2(2{\rm tan}^{-1}e^x)\nonumber\\ +\frac{1}{2}{\rm cos}^2({\rm
tan}^{-1}e^x)+ \frac{\epsilon}{m^*\beta}{\rm cos}^2(2{\rm tan}^{-1}e^x)\big]   \ \
. \end{eqnarray} Equation (12) represents the single-body Hamiltonian for a
ficticious  particle of energy  $\epsilon$ and unit mass, moving in the xy-plane
in a non-linear (i.e. $\epsilon$-dependent)  potential  $ V(x,y;\epsilon)$, 
subject to a ficticious, non-uniform magnetic field  in the  $z$-direction
\begin{equation} B_{\rm fic}= \frac{\hbar J c}{4e}\big[1-{\rm cos} (4({\rm
tan}^{-1}e^x))\big]\ \ . \end{equation}  $B_{\rm fic}$ is independent of $B$ and
has a maximum of $\frac{\hbar |J| c}{2e}$   at $x=0$ for all $y$.  For small $x$,
$B_{\rm fic}\approx \frac{\hbar J c}{2e} (1-x^2)$. As  $x\rightarrow\pm\infty$,
$B_{\rm fic}\rightarrow 0$.  Note we have here chosen to highlight the  
Schrodinger-like form of Eq. (12); a simple rearrangement of Eq. (12) 
shows it to be hermitian with a weighting function  ${\rm sin}^2(2{ \rm
tan}^{-1}e^x)$.   These results are exact so far. 
Figure 3 shows the potential $ V(x,y;\epsilon)$ in  the $(x,y)$ plane.
$V(x,y;\epsilon)\geq 0$ everywhere.  Minima occur at $(0,0)$ and $(0,\pm\pi)$ where
$V(x,y;\epsilon)=0$ (N.B. $(0,\pi)$ is equivalent to  $(0,-\pi)$).  Maxima occur
at $({\rm ln} {\sqrt 3}, \pm\frac{\pi}{2})$  in Fig. 3,  where
$V(x,y;\epsilon)\rightarrow\infty$.  Since $\epsilon\geq 0$, these statements hold
for any  $\epsilon$.  We now discuss the physical significance of these  features. 
The classical configurations of minimum energy   (Wigner molecule) correspond to
the  particles  lying on a ring in the form of an equilateral triangle with  
$V_{\rm class}=\omega_0(B)[6m^*\beta]^{\frac{1}{2}}$.  There are two distinct
configurations, i.e. two distinct symmetrically equivalent minima
\cite{term}, with clockwise  orderings  $\Omega'_0\equiv(132)$ and 
$\Omega'_1\equiv(123)$ corresponding to $(\alpha,\theta)=(\frac{\pi}{4},\pm
\frac{\pi}{2})$. 
 In
$(x,y)$  coordinates, these  correspond to $(0,0)$ and $(0,\pi)$ (equivalently, 
$(0,-\pi)$).  Hence the classical configurations coincide with the  minima in $
V(x,y;\epsilon)$ in Fig. 3 and the maximum in  $B_{\rm fic}$.  As pointed out in
Ref. 12, the formation of a Wigner molecule should therefore be 
favored by  both large $B_{\rm fic}$ (i.e. large $|J|$) and deep  $
V(x,y;\epsilon)$  minima  (i.e. large $\beta$, strong electron-electron 
interactions).

Consider the limit of very  strong  electron-electron 
interactions  (i.e. $\beta\rightarrow \infty$).  Since the tunnel barrier height
between the two  $ V(x,y;\epsilon)$ minima $\sim \beta$, the ficticious  particle 
sits at one of these minima and the system is locked  in one of the two classical
configurations, e.g. $\Omega'_0\equiv(132)$  at    $(0,0)$.  The tunnelling
probability between the minima $\Omega'_0$ and $\Omega'_1$ is zero.  Tunnelling 
between the  two minima implies a mixture of configuration $(123)$ into $(132)$
and hence interchange of the original  electrons; in  many-body  language exchange
effects arising from wavefunction  antisymmetry  are therefore negligible.  
$\epsilon$ is small compared to $m^*\beta$ and Eq. (12)  reduces to
\begin{equation} E_{\rm rel}=\hbar\omega_0(B)\big[2n+
(1+J^2+\frac{6m^*\beta}{\hbar^2})^{\frac{1}{2}} +1\big]+J\frac{\hbar\omega_c}{2}\
\ . \end{equation} The energy $E_{\rm rel}\geq V_{\rm class}$ since it includes
the hyperradial  zero-point energy  (N.B. $\hbar\rightarrow 0$  yields $E_{\rm
rel}\rightarrow V_{\rm class}$ and  $B_{\rm fic}\rightarrow 0$). 

Next consider large but finite $\beta$.  The ficticious particle now    moves in
the vicinity of the minimum $\Omega'_0$ (i.e.  $(x,y)\approx (0,0)$).  The
electrons in the Wigner solid are effectively  vibrating around  their classical
positions. Expanding the potential $ V(x,y;\epsilon)$ about $(0,0)$  to third 
order,  the exact Eq. (12) becomes  \begin{equation}
\big[\frac{p_x^2}{2}+\frac{(p_y-\frac{\hbar J x}{2})^2}{2} \\
+\frac{1}{2}\omega_x^2x^2+\frac{1}{2}\omega_y^2y^2\big]  G(x,y)=\epsilon G(x,y)
\end{equation} where $\omega_x^2=(\frac{3m^*\beta}{4}+2\epsilon)$  and    
$\omega_y^2=\frac{3m^*\beta}{4}$. This has the form  of a single electron moving
in an anisotropic parabolic  potential,  subject to a uniform  magnetic field
$B_{\rm fic}=\frac{\hbar J c}{2e}$. Equation (16) is exactly solvable for
$\epsilon$  using a symmetric  gauge\cite{madhav} (the energies are independent 
of the choice of  gauge for $B_{\rm fic}$). As an illustration,  we consider 
small $\epsilon$ hence $\omega_x\approx\omega_y$.  The relative energy is
then given by  \begin{eqnarray} E_{\rm rel}=\hbar\omega_0(B)(2n+
\big[1+J^2+\frac{6m^*\beta}{\hbar^2}+ 2(2n'+|l|'+1)
(J^2+\frac{12m^*\beta}{\hbar^2})^{\frac{1}{2}}+2l'J \big]^{\frac{1}{2}}
+1)\nonumber\\ +J\frac{\hbar\omega_c}{2}\ \ . \end{eqnarray} The ficticious
particle has its own set of Fock-Darwin (and hence Landau) levels \cite{fock} 
labelled by  $n'$ and a  ficticious angular momentum $l'$.  For large $\beta$ and
small $n',l'$ and $J$, Eq. (17)  yields an  oscillator excitation spectrum with
two characteristic frequencies ${\sqrt 2}\hbar\omega_0(B)$ and $2\hbar\omega_0(B)$ 
representing  shear and  breathing modes of the Wigner molecule. 

For smaller $\beta$ (i.e. weaker interactions) and/or larger  $\epsilon$ (i.e. 
excited states),  the tunneling probability between the $V(x,y;\epsilon)$ minima 
$\Omega'_0$ and $\Omega'_1$ in Fig. 3 becomes significant. The Wigner molecule 
begins to melt and wavefunction antisymmetry must be included. This is
discussed further in Sec. III. As mentioned in Ref. 12, the resulting
analytically obtained magic-number $J$ transitions are found to be in good
agreement with the numerical results for $1/r$ interaction. We note that the
analytic results become more  accurate in the  Wigner solid regime (e.g. large
$\beta$ or $|J|$) while the  numerical calculations become more computationally
demanding.

\vskip 0.2in {\bf II.3 Simplified hyperangular equations for arbitrarily
large $N$}

For general $N$ the hyperrangular equation (Eq. (6))  is 
$(2N-4)$-dimensional.  However in the Wigner solid regime (large $\beta$ or $|J|$) 
the classical minimum energy configurations will still be  important in
determining $\epsilon$ and hence $E_{\rm rel}$, just as for $N=3$.  Here we will
consider the limit that the number of electrons is large ($N>>1$). This is the
limit of interest in the FQHE and in large quantum dots. Specifically,
we will introduce in this section a series of approximations in order
to simplify the exact hyperangular equation. At each stage, the
corresponding simplified hyperangular equation is explicitly given. The resulting
discussion is detailed -- however we feel that this is
necessary in order to justify the successively simpler (and more
approximate) hyperangular equations. Each of these simplified
hyperangular equations can be solved numerically: the complexity of the
algorithms needed obviously decreases as more approximations
are introduced. However, the goal in this paper is to obtain a simplified version
of the hyperangular equation which can be treated analytically, but which is
still based on a set of reasonable approximations. 

Figure 4 shows the classical ground-state configuration for $N=230$ electrons
(black dots) in a parabolic quantum dot, as obtained by Bedanov and Peeters using
a Monte Carlo algorithm \cite{Bedanov}. The rings are drawn as a guide to
the eye. The number
$N$ of electrons is relatively small in the context of the $N\rightarrow\infty$
limit and hence the details of the ground-state configuration, particularly for
larger rings, will be prone to edge effects. However, the inner rings show a nearly
hexagonal lattice as expected for the $N\rightarrow\infty$ limit. For the purposes
of illustration we will therefore consider Fig. 4 as being representative of the
$N\rightarrow\infty$ classical configuration.  Consider the {\em particular}
classical configuration
$\Omega'_0$ where the $N$'th electron is near the center and the first electron is
on the circumference of the droplet. This is shown schematically in Fig. 5. As
discussed in Sec. II.2 for
$N=3$, the fully quantum mechanical system will also lie near this configuration in
$\Omega'$-space in the limit of very large $\beta$.  The $j$'th Jacobi coordinate 
is given by  \begin{equation} {\bf
X}_j=\big[{\frac{j-1}{j}}\big]^{\frac{1}{2}}({\bf r}_j-{\bf R}_{j-1})
\end{equation}  where ${\bf R}_{j-1}=\frac{1}{j-1}({\bf r}_1+ {\bf r}_2+\dots {\bf
r}_{j-1})$: this quantity ${\bf R}_{j-1}$ can be thought of as the
`center-of-mass' of the electrons $1,2\dots {j-1}$. For $j>>1$, and for
configurations as in Figs. 4 and 5 where the electrons are evenly distributed around
the origin, the quantity ${\bf R}_{j-1}$  will be small compared to the typical
electron lattice spacing. In addition, the prefactor
$[{\frac{j-1}{j}}]^{\frac{1}{2}}\rightarrow 1$ for large $j$. Hence ${\bf
X}_j\rightarrow {\bf r}_j$ for large $j$. However, there is an exact identity for
hyperangular Jacobi coordinates: $\sum_{j=2}^N [\frac{X_j}{r}]^2=1$. Given that
$X_j\rightarrow 0$ as $j\rightarrow N$ for large $N$, this implies that each term
$[\frac{X_j}{r}]^2<<1$ for large $j$. Given the definition of the hyperangular
coordinates stated earlier on, it follows that the hyperangles $\alpha_j<<1$ for
$j>>1$. Hence to first order in $\alpha_j$,  we can approximate $X_N=r{\rm sin}
\alpha_N\approx r\alpha_N$. Similarly $X_{N-1}=r{\rm cos} \alpha_N {\rm sin}
\alpha_{N-1}\approx r\alpha_{N-1}$ and, more generally,  $X_j\approx r\alpha_{j}$.
To summarize, for configurations similar to that shown in Fig. 5, we have the
approximate result ${\bf X}_j\approx r\alpha_je^{i\theta_j}$ in the $N>>1$ and
$j>>1$ limit. There are two points to note: although we need both $j>>1$ and
$N>>1$, $j$ can still be an order of magnitude less than $N$. Second, the error
introduced by assuming ${\rm sin}\alpha_j\approx \alpha_j$ is still reasonably
small even for $j=2$ (recall $\alpha_2=\frac{\pi}{2}\approx 1.57$ as compared to
${\rm sin}\alpha_2=1$. To remain consistent within our approximation, we will take
$\alpha_2=1$ instead of $\frac{\pi}{2}$ in what follows).   

This approximate form for ${\bf X}_j$ leads to an interesting simplification of
the exact hyperangular equation. The small-angle (i.e.
$\alpha_j<<1$) limit of Eq. (6) yields:  \begin{equation}
\big[\sum_{j=2}^N -\frac{\hbar^2}{2m^*}\nabla^2_j + \beta V(\Omega)\big] F(\Omega)=
\frac{\hbar^2}{2m^*}\gamma(\gamma+2N-4) F(\Omega) \end{equation}
where  \begin{equation} \nabla^2_j\equiv
-\frac{\partial^2}{\partial\alpha_j^2}-\frac{1}{\alpha_j}
\frac{\partial}{\partial\alpha_j}-\frac{1}{\alpha_j^2}
\frac{\partial^2}{\partial\theta_j^2} \end{equation} is the two-dimensional
Laplacian for a ficticious particle with position $(\alpha_j,\theta_j)$ in polar
coordinates, the potential energy term
\begin{equation}
V(\Omega)\equiv
V(\Omega')\sim\sum_{j<j'}|{\alpha}_je^{i\theta_j}-
{\alpha}_{j'}e^{i\theta_{j'}}|^{-2}
\end{equation} and $F(\Omega)=e^{iJ\theta'}G(\Omega')$.
This equation is a good approximation for $j\rightarrow N$ with $N>>1$ but becomes
worse as $j\rightarrow 0$ and/or $N\rightarrow 0$. (Recall $\alpha_2=1$ hence the
sum can start from $j=2$ as shown). However this is sufficient for the purposes of
this paper since we are interested in states that evolve within the bulk of the
$N$-electron droplet  as opposed to those at the edge. Physically, Eq. (19)
describes a set of $N-1$ ficticious particles moving on a two-dimensional plane
subject to a two-body inverse-square interaction, in the {\em absence} of a
magnetic field. It is interesting to note this transformation of having replaced an
$N$-particle problem in a magnetic field with an $N-1$ particle problem without a
magnetic field seems reminiscent of composite fermion constructions at
half-integer filling fractions. The effective Schrodinger equation in Eq.
(19) carries the following constraint: the exact hyperangular identity
$\sum_{j=2}^N [\frac{X_j}{r}]^2=1$ implies $\sum \alpha_j^2\sim 1$. This may
complicate any attempt at a solution using a `renormalization' type of approach,
such as the setting up of a recursion equation relating $\gamma$ for $N$ particles
to $\gamma$ for $N-1$.

It is more useful to view Eq. (19) in the context of
a single ficticious particle moving in a multi-dimensional space containing
potential miminima corresponding to the various classical minimum energy
configurations $\{\Omega'_i\}$. This directly connects the $N$-electron problem to
the $N=3$ problem discussed in Sec II.2. As discussed in Sec. II.1, a Jacobi
transformation can be undertaken on the $\{\theta_j\}$ variables. In particular,
\begin{equation}
{\theta}_{[j]}=\big[{\frac{j-2}{j-1}}\big]^{\frac{1}{2}}({\theta}_j-{\Theta}_{j-1})
\end{equation}    where $j=3,4\dots N$ and
${\Theta}_{j-1}=\frac{1}{j-2}({\theta}_2+ {\theta}_3+\dots {\theta}_{j-1})$. The
quantity ${\Theta}_{j-1}$ represents the average of the angles $\theta_j$ where
$j=2,3\dots {j-1}$. For $j>>1$ the quantity ${\Theta}_{j-1}$  will be
approximately a constant, $\bar\Theta$, since the $j-1$ particles are evenly distributed
about the origin in a given $\Omega_i'$ configuration (recall Figs. 4 and 5). In addition,
the prefactor $\big[{\frac{j-2}{j-1}}\big]^{\frac{1}{2}}\rightarrow 1$ for large
$j$. Hence  ${\theta}_{[j]}\rightarrow {\theta}_j-\bar\Theta$ for large $j$, neglecting
terms of order $(\frac{1}{N})$. With $F(\Omega)=e^{iJ\theta'}G(\Omega')$, Eq. (19)
further reduces to 
$$\big[\sum_{j=3}^N -\frac{\hbar^2}{2m^*}
(\frac{\partial^2}{\partial\alpha_j^2}+\frac{1}{\alpha_j}
\frac{\partial}{\partial\alpha_j}+\frac{1}{\alpha_j^2}
[\frac{\partial}{\partial\theta_{[j]}}+\frac{iJ}{N-1}]^2)
 + \beta V(\Omega')\big] G(\Omega') =$$
\begin{equation} 
\frac{\hbar^2}{2m^*}\gamma(\gamma+2N-4)G(\Omega') \end{equation} Again this
equation is a good approximation for $j\rightarrow N$ but becomes worse as
$j\rightarrow 0$. The $\{\theta_{[j]};\alpha_j\}$
manifold carries the following constraints:  $\sum\alpha_j^2\sim 1$ and
$\sum\frac{\partial}{\partial\theta_{[j]}}\sim 0$. The latter condition is an
approximate identity for large $N$ and is obtained by combining $\sum
\frac{\partial}{\partial \theta_j}=\frac{\partial}{\partial \theta'}$ (this is an
exact property of any Jacobi transformation)  and
$\frac{\partial}{\partial \theta_j}\sim \frac{\partial}{\partial \theta_{[j]}}+
\frac{1}{N-1}\frac{\partial}{\partial \theta'}$. This new condition hence 
reflects the fact that the total relative angular momentum is only associated
with the $\theta'$ variable: there is no additional contribution to the
relative angular momentum contained within the $\Omega'$ dynamics. These
approximate constraints allow us to make a further simplification of the
hyperangular equation as follows. Using the  approximate identity
$\sum\alpha_j^2\sim 1$ we can define an average hyperangle $\bar\alpha\sim
N^{-\frac{1}{2}}$. We will therefore replace the term
$\sum\frac{1}{\alpha_j^2}\frac{J^2}{(N-1)^2}$ in Eq. (23) by 
$\frac{1}{\bar\alpha^2}\sum\frac{J^2}{(N-1)^2}\sim J^2$ assuming large $N$. 
We
can hence rewrite Eq. (23) in the form 
$$ \big[\sum_{j=3}^N -\frac{\hbar^2}{2m^*}
(\frac{\partial^2}{\partial\alpha_j^2}+\frac{1}{\alpha_j}
\frac{\partial}{\partial\alpha_j}+\frac{1}{\alpha_j^2}
[\frac{\partial^2}{\partial\theta_{[j]}^2}+\frac{2iJ}{N-1}
\frac{\partial}{\partial\theta_{[j]}}])
 + \beta [V(\Omega')-V(\Omega'_0)]\big] G(\Omega')= $$
\begin{equation}
\frac{\hbar^2}{2m^*}
[\gamma(\gamma+2N-4)-J^2-\frac{2m^*\beta}{\hbar^2}V(\Omega'_0)]
G(\Omega')\ . \end{equation} 
Although its derivation has involved approximations, Eq. (24) merits some
discussion since it elucidates several of the statements made in Sec. II.1. The
right-hand side is just $\frac{4}{m^*}\epsilon$. Given that
$\sum\alpha_j^2\sim 1$ and
$\sum\frac{\partial}{\partial\theta_{[j]}}\sim 0$, the $J$-dependence of $\epsilon$
will tend to be weaker than $J^2$ as claimed earlier. Note that since 
$\sum\alpha_j^2\sim 1$, the moment of inertia $I$ of a given classical
configuration in $\Omega$-space, treated as a rigid body, is just $m^*$. Hence the
classical rigid-body rotational energy $\frac{\hbar^2 J^2}{2I}\sim \frac{\hbar^2
J^2}{2m^*}$ which is precisely the term appearing in the right-hand side of Eq.
(24). This then justifies the statement made in Sec. II.1 that
$\epsilon$ excludes the classical rigid-body rotational energy.
The term $V(\Omega'_0)$ denotes $V(\Omega')$ evaluated at a given classical
SEM equilibrium configuration $\Omega'\equiv \Omega'_0$. We emphasize that
$V(\Omega'_0)\equiv  V(\Omega'_i)$, i.e. same potential energy for all SEMs.
Since $\Omega'_0$ is a minimum, the difference term  $[V(\Omega')-V(\Omega'_0)]$
can be expanded around $\Omega'_0$. The
leading terms will be quadratic in $\theta_{[j]}-\theta_{[j0]}$ etc. Hence
$\epsilon$ does indeed describe the zero-point energy associated with the spread
in $G(\Omega')$ around the classical minima, as claimed in Sec. II.1
and shown explicitly for $N=3$ in Sec. II.2. This point is further discussed
below for large $N$. 

The hyperangular equation Eq. (24) is now simpler, however it is still not quite in
a form which makes it amenable to analytic calculation. This final step can be
achieved with the following considerations. Given the two approximate constraints
$\sum\alpha_j^2\sim 1$ and
$\sum\frac{\partial}{\partial\theta_{[j]}}\sim 0$, the term involving $\sum
\frac{1}{\alpha_j^2}{\frac{\partial}{\partial\theta_{[j]}}}$ should be
small as compared to the term involving $\frac{1}{\alpha_j^2}
\frac{\partial^2}{\partial\theta_{[j]}^2}$, and hence will be neglected.
Furthermore just as for $N=3$, we are initially considering  the quantum
mechanical solution near a given classical minimum $\Omega'_0$, i.e.
$\beta\rightarrow\infty$. Hence the term $\frac{1}{\alpha_j^2}
\frac{\partial^2}{\partial\theta_{[j]}^2}$ can be approximated by 
$\frac{1}{\alpha_{j0}^2} \frac{\partial^2}{\partial\theta_{[j]}^2}$ where
$\alpha_{j0}$ is the value of $\alpha_j$ at $\Omega'\equiv\Omega'_0$. The fact
that $\Omega'_0$ is a minimum suggests that the leading order expansion of 
$[V(\Omega')-V(\Omega'_0)]$ will involve terms like
$(\theta_{[j]}-\theta_{[j0]})^2$ and $(\alpha_{j}-\alpha_{j0})^2$ for all $j$
but not cross terms: this was demonstrated explicitly for $N=3$ earlier
where $V(x,y)$ was found to be a function of $x^2$ and $y^2$ but not $xy$.  This
implies the following simplification  for $\Omega'\sim\Omega'_0$: 
$[V(\Omega')-V(\Omega'_0)]\sim \sum_j [v(\alpha_j-\alpha_{j0})+w(\theta_{[j]}-
\theta_{[j0]})]$ where both $v$ and $w$ have a minimum at $\Omega'\equiv
\Omega'_0$, i.e. at $\alpha_j=\alpha_{j0}$ and $\theta_{[j]}=\theta_{[j0]}$. The
approximate separability of the potential suggests that the hyperangular function
$G(\Omega')$ can now be written as $f(\{\alpha_j-\alpha_{j0}\})
g(\{\theta_{[j]}-\theta_{[j0]}\})$ where  the functions $f$ and $g$ are peaked
around $\Omega'\equiv \Omega'_0$.  This was shown to be true explicitly for $N=3$
where $f$ and $g$ turned out to be gaussians (harmonic oscillator wavefunctions).
Since $\Omega'_0$ is still a minimum point for large $N$, $f$ and $g$ will retain
their gaussian-like character for general $N$.  We will therefore write 
$g(\{\theta_{[j]}-\theta_{[j0]}\})\sim \prod_j g_j(\theta_{[j]}- \theta_{[j0]})$
where $g_j(\theta_{[j]}-\theta_{[j0]})$ is a function peaked around the minimum
coordinate  $\theta_{[j]}=\theta_{[j0]}$. The hyperangular equation is now fully
separable into an equation involving $\{\alpha_j\}$ \begin{equation}
\big[\sum_{j=3}^N -\frac{\hbar^2}{2m^*}
(\frac{\partial^2}{\partial\alpha_j^2}+\frac{1}{\alpha_j}
\frac{\partial}{\partial\alpha_j})
 + \beta v_j(\alpha_j-\alpha_{j0})\big] f(\{\alpha_j-\alpha_{j0}\})= 
E_\alpha f(\{\alpha_j-\alpha_{j0}\})
\end{equation}
together with the following equations for each $\theta_{[j]}$:
\begin{equation}
\big[-\frac{\hbar^2}{2m^*\alpha_{j0}^2}\frac{\partial^2}{\partial\theta_{[j]}^2}
+\beta w_j(\theta_{[j]}-\theta_{[j0]})\big]g_j(\theta_{[j]}-\theta_{[j0]})=
e_jg_j(\theta_{[j]}-\theta_{[j0]})\ \ .
\end{equation}
The relation between $\epsilon$, $E_\alpha$ and $e_j$ is as follows:
$\frac{4}{m^*}\epsilon =E_\theta+E_\alpha$
where $E_\theta=\sum_j e_j$. The full
expression for the relative energy hence becomes 
\begin{equation} E_{\rm rel}=\hbar\omega_0(B)\big[2n+
\big([N-2]^2+J^2+\frac{2m^*\beta}{\hbar^2}V(\Omega'_0)
+ {\frac{2m^*}{\hbar^2}}[E_\theta+E_\alpha]
\big)^{\frac{1}{2}} +1\big]\\ 
+J\frac{\hbar\omega_c}{2}\ \ .  \end{equation} 
Since Eqs. (25) and (26) have a Schrodinger-like form with $E_\alpha$ and
$e_j$ as eigenvalues respectively, we will refer to these two quantities as
`energies' even though this is not strictly correct terminology.

\vskip 0.2in

{\bf II.4 Characteristics of the low-energy solutions $G(\Omega')$}

So far we have considered the solutions near a particular minimum
configuration $\Omega'_0$, i.e. we have considered very large $\beta$ just as we
did initially for $N=3$.  Very large $\beta$ implies that $G(\Omega')$ will be
peaked around one of the SEMs, e.g. around $\Omega'_0$.
In the limit of zero tunnelling between SEMs, the
solution $G(\Omega')\sim f(\{\alpha_j-\alpha_{j0}\})
g(\{\theta_{[j]}-\theta_{[j0]}\})$ centered around $\Omega'_0$ will be {\em
degenerate} with the identically localized solutions centered at all other
SEMs $\{\Omega'_i\}$. These localized functions can be
thought of as  atomic-like orbitals in $\Omega'$ space. In particular there will
be a set of orbitals associated with each SEM $\Omega'_i$.
The corresponding coordinates and hyperangular equations describing these solutions
are identical to those obtained earlier in Sec. II.3: however the spatial
ordering of the electrons for the various $\Omega'_i$ minima will necessarily
change; for example electron $N$ will not necessarily be close to the center.
Using the usual variational argument for Schrodinger-like equations, the lowest
energy (i.e. lowest $E_\alpha$ and $E_\theta$) solutions of Eqs. (26) and (27)
will be those with the minimum number of nodes.  

For large but finite $\beta$
there will be a small but finite tunnelling between the various minima 
$\{\Omega'_i\}$, hence the
complete solution $G(\Omega')$ will be more correctly described as a linear
combination of the atomic-like solutions, just as in a single-particle
tight-binding model. 
Furthermore for $N\geq 6$, as noted earlier, there will be additional classical
minima which are not topologically equivalent; again borrowing from the language of
molecular physics \cite{Eckardt,term} these minima are termed {\em symmetrically
inequivalent minima} (SIM). These SIMs are local minima in $\Omega'$-space which
are often just slightly higher in energy than the SEMs $\{\Omega_i\}$. In the large
$N$ limit, these minima correspond to defect states in a
hexagonal crystal. Fisher, Halperin and Morf \cite{Fisher} showed that a Wigner
crystal with a localized defect (WXD) can be quite close in energy to the perfect
Wigner crystal (WX). This finding was recently verified in the context of $N$
electrons in a two-dimensional parabolic quantum dot by Bolton \cite{bolton} and
Bedanov et al. 
\cite{Bedanov}. These authors all found that the global minimum for the classical
$N$-electron system tends towards a hexagonal crystal as
$N\rightarrow\infty$, as expected for the Wigner crystal (WX). However
configurations corresponding to a Wigner crystal with single defects (WXD) are
only slightly higher in energy. In the language of the present paper the WX
represents the SEMs while the WXD represents the SIMs. Although the SIMs are not
true global minima, the complete solution $G(\Omega')$ should certainly include
finite mixing with them. This is particularly true since the `nearest-neighbors'
of a given SEM in $\Omega'$-space are SIMs. This is simply a consequence of the fact
that translation between two adjacent SEMs in $\Omega'$-space requires interchange
of at least two electrons, while translation between a given SEM and its nearest
SIMs requires only slight electron distortion. Each SEM minimum $\Omega_i$ will
have $p$ defect states as its nearest neighbors in $\Omega'$-space -- we denote
these nearby SIM minima as $\{\Omega'_{i;a}\}$ where $a=1,2\dots p$ (c.f. Fig. 2). 

The resulting wavefunction $G(\Omega')$ will therefore
resemble a tight-binding LCMO (Linear Combination of Molecular Orbitals)
wavefunction where each `molecule' consists of `atomic' orbitals on one of the SEM
minima $\Omega_i$ mixed with `atomic' orbitals on each of its nearby SIM minima.
We emphasize that $N=3$ has no SIM minima. $N=6$ is the smallest $N$ having SIMs.
The SIMs for $N=6$ consist of a six-member ring configuration while the SEM's
contain a five-member ring plus one electron at the center \cite{bolton,Bedanov}.
For general
$N$, the low-energy solutions should therefore be reasonably well-described by
\begin{equation} G(\Omega') \sim \sum_i^{SEM} \sum_a^{SIM} c_{i;a} 
f(\{\alpha_j-\alpha_{ji;a}\}) g(\{\theta_{[j]}-\theta_{[ji;a]}\})\ \ .
\end{equation}
It is well-known from elementary tight-binding theory that the lowest-energy
states are `bonding' wavefunctions of $s$-orbitals. In the present context, we
similarly expect the lowest energy $G(\Omega')$ to have as few nodes as
possible (i.e. it will be gaussian-like around each of the SEM $\{\Omega'_i\}$
thereby resembling an $s$-orbital); it will also correspond to the coefficients
$c_{i;a}$ being identical for each $i$ (i.e. it will resemble a `bonding' state).

\vskip 0.2in

{\bf III. FERMION STATISTICS, MAGIC NUMBERS AND FILLING FRACTIONS}

So far we have not introduced the requirement that the total $N$-electron
wavefunction be antisymmetric. In this section, we will show that
it is precisely this requirement that produces the observed FQHE filling factors
for large $N$.
 
It is useful to first discuss the effect of antisymmetry in the case of $N=3$
electrons before considering large $N$. For  three  spin-polarized electrons,
$\psi$ must be antisymmetric under particle interchange  $i\leftrightarrow j$. 
The hyperradial part $R(r)$ is  invariant; particle permutation operations in 
$({\bf r}_1,{\bf r}_2,{\bf r}_3)$  become straightforward {\em space-group}
operations in the  $(x,y)$ plane.  For small $(x,y)$, $1\leftrightarrow 2$ is
equivalent to $(x,y)\rightarrow (x,y+\pi)$ with  $\theta'\rightarrow \theta'+
\frac{\pi}{2}$; $1\leftrightarrow 3$ is equivalent to $(x,y)\rightarrow (\bar
x,\bar y-\pi)$  with $\theta'\rightarrow \theta'+ \frac{\pi}{6}$ ($(\bar x, \bar
y)$ represents $(x,y)$  rotated by  $\frac{4\pi}{3}$); $2\leftrightarrow 3$ is
equivalent to $(x,y)\rightarrow (\tilde x,\tilde y+\pi)$  with $\theta'\rightarrow
\theta'- \frac{\pi}{6}$ ($(\tilde x, \tilde y)$ represents $(x,y)$  rotated  by 
$-\frac{4\pi}{3}$).  The solutions $G(x,y)$ of
Eq. (12) with the lowest possible  $\epsilon$  and hence lowest $E_{\rm rel}$ at a
given $\omega_c$, should be  nodeless  in the vicinity of $(0,0)$  (c.f. ground
state in the parabolic potential with $n'=0=l'$ in  Eq. (17)). However the above
symmetry requirements forbid such a nodeless  solution unless
$e^{i\pi\frac{2J}{3}}=1$. 
Therefore the only symmetry-allowed  solutions
$G(x,y)$ which are nodeless are those where $J$ is a multiple of three, as
observed in numerical calculations for $N=3$ electrons with a $1/r$ interaction. 
It is important to note that this condition, i.e. $e^{i\pi\frac{2J}{3}}=1$,
just arises
from combining the effect of any two sets of particle interchanges
$i\leftrightarrow j$. For $N=3$, two sets of particle interchanges
correspond to rotations of a given SEM: this can be seen simply as follows.
Consider a given SEM in Fig. 3, e.g. $\Omega'_0\equiv(132)$. Interchanging
$1\leftrightarrow 2$ and $2\leftrightarrow 3$ yields the {\em same} SEM, i.e. 
(132), rotated anticlockwise by $\frac{2\pi}{3}$. Hence combinations of 
two sets of particle interchanges merely rotate the Wigner molecule without
involving a transformation from one SEM to another, i.e. without moving from (132)
to (123).
   Hence in order to obtain the `magic'
angular momentum values for $N=3$, it is {\em sufficient} to consider the subset of
particle interchanges from the $S_3$ permutation group which correspond to
point-group rotations $C_3$, i.e. those which do not correspond to translations
between SEMs. This result is discussed by Maksym in Ref.
15 following earlier work on molecules by Wilson \cite{term}. Maksym also argued for
$N=3$ that the remaining permutations which correspond to translations between
SEMs, and hence the effect of tunnelling between SEMs, represents a small
perturbation which does not affect the magic $J$-values. In contrast, for $N=6$
where topologically distinct classical configurations coexist, the tunnelling
between SEMs and SIMs plays a crucial role in determining the $J$ values
of the low-energy ground states of the system. In particular, Maksym pointed out
that tunnelling between SEMs and SIMs should be most favorable when the SEM and
SIM configurations have a common $J$ value. This is consistent with
analogous ideas in single-particle tight-binding theory, where the overlap matrix
element (and hence bandwidth) is larger between $s$ orbitals than between $s$ and
$p$ orbitals. Maksym conjectured that the resulting tunnelling might lead to
`liquid'-like states with a lower overall energy. 

These considerations motivate us to follow a similar strategy for $N$
electrons. In particular, we will show that considering just a subset of particle
interchanges of $S_N$ corresponding to rotations of rings within the Wigner molecule
(WX) and Wigner molecule plus defect (WXD) is {\em sufficient} to determine the
magic
$J$-values corresponding to the observed FQHE filling factors.
As for $N=3$, we focus on the vicinity of a given SEM, e.g.
$\Omega'_0$.  
Following the discussion in Sec. II.3,
interchanging ${\bf r}_i\leftrightarrow {\bf r}_j$ is relatively
straightforward for $i,j>>1$ since ${\bf X}_j\approx r\alpha_je^{i\theta_j}\sim
{\bf r}_j$. Neglecting terms of order $(\frac{1}{N})$, it just corresponds to
${\alpha}_i\leftrightarrow {\alpha}_j$ and  ${\theta}_i\leftrightarrow
{\theta}_j$. The derivation of the
transformation rules including terms of order $(\frac{1}{N})$, is
straightforward but tedious. Just as for $N=3$, however, it turns out in what
follows that we do not need to consider individual $i\leftrightarrow j$
transformations. 

\vskip 0.2in

{\bf III.1 Spin-polarized system}

Consider the classical configuration $\Omega'_0$ shown in Figs. 4 and 5. For
large $j$ the electrons can be thought of as forming an approximately ring-like
structure. Counting the number of rings from the center outwards, the first ring
contains 6 electrons, the second contains approximately 12 and so on. We first
focus on a `typical' ring without any defects: it will contain a large, even
number $N_m$ of electrons (approximately  $6m$ electrons where $6m>>1$) but these
electrons will have an index $j>>1$, i.e. we are not considering rings near the
edge of the $N$-electron droplet. We are going to consider just the subset of all 
particle interchanges $i\leftrightarrow j$  which are equivalent to rotations of
this $m$'th ring. Since ${\bf X}_j\approx r\alpha_je^{i\theta_j}\sim {\bf r}_j$,
all members of the ring have approximately the same $\alpha$, i.e.
$\{\alpha_j\}\equiv\alpha_m$ for all $j$ in ring $m$. Hence interchanging two
members of the ring just involves a transformation between their $\theta_j$
coordinates. Since all members of the ring have a similar environment and the same
$\alpha_j$, the potential energy $w_j$ and hence $g_j$ in Eq. (26)  will have the
same form for all $j$ in the ring $m$. As for $N=3$, the lowest-energy solutions
should be those with $g_j$ nodeless; $g_j$ will be an approximately gaussian
function of $\theta_{[j]}$ centered around $\theta_{[j0]}$. Hence we can write
$g_j\equiv g_m$ for all $j$ in ring $m$.  We now rotate the electrons in the ring,
and hence the ring itself, by an angle $\frac{2\pi}{N_m}$. Since ${\bf
X}_j\sim{\bf r}_j$ this corresponds to
$\theta_j\rightarrow\theta_j+\frac{2\pi}{N_m}$. The $g_m$ functions are nodeless
and (in a given ring) identical, hence the product $\prod_j
g_j(\theta_{[j]}-\theta_{[j0]})$ for $j$ in ring $m$ can be replaced by  $\prod_j
g_m(\theta_{[j]}-\theta_{[j0]})$.  The transformation keeps the system within
the subset of all SEMs corresponding to the same ring ordering, i.e. just as
for $N=3$ the rotation operation in real space becomes a space-group operation in
$\Omega'$-space which translates the system between SEMs.
Recall that $G(\Omega')$ for minimum-energy states should resemble a `bonding'
linear-combination of $s$-like orbitals  (i.e. an approximately gaussian
$\theta$-dependence around the various SEMs (Eq. (28))).  The coefficients
$c_{i;a}$ in the expression for $G(\Omega')$ in Eq. (28) will therefore be
identical for all the SEM minima $\{\Omega'_i\}$ corresponding to this same ring
ordering. Since $G(\Omega')$ corresponds to a linear combination of identical
orbitals with the same coefficient, the overall effect of the transformation on 
$G(\Omega')$ due to ring rotation will be quite small; we simply move between this
subset of SEMs $\{\Omega'_i\}$ each of which has the same local orbitals. In
contrast, the effect on the $\theta'$ variable is relatively important; it follows
from Eq.  (8) that  $\theta_j\rightarrow\theta_j+\frac{2\pi}{N_m}$ corresponds to 
$\theta'\rightarrow\theta'+ \Delta\theta'$ where
$\Delta\theta'=N_m\frac{2\pi}{N_m(N-1)}= \frac{2\pi}{(N-1)}$. The total function
$F(\Omega)=e^{iJ\theta'}G(\Omega')$ hence becomes  $e^{iJ\Delta\theta'}F(\Omega)$.
Given that $N_m$ is an even number, rotation of the $m$'th ring by
$\frac{2\pi}{N_m}$ necessarily corresponds to an odd number of interchanges 
$i\leftrightarrow j$. If we assume the electrons are {\em spin polarized}, the
spatial part of the $N$-electron wavefunction must be totally antisymmetric and
hence the overall phase change must equal $e^{i\pi(2n+1)}$ where $n$ is any
integer. Denoting the $J$ value as $J_{WX}$ we therefore obtain the condition
\begin{equation} J_{WX}=\frac{1}{2}(N-1)(2n+1) \ . \end{equation}  Importantly,
this criterion for $J_{WX}$ is {\em independent} of $m$ and hence holds for all
rings $m$. In other words, this criterion guarantees that the $N$-electron
wavefunction  has the correct permutational symmetry under the subset of all
permutations of $N$ electrons which  correspond to ring rotations. Note that
$J_{WX}$ must be an integer. 

Now consider the Wigner crystal plus defect (WXD). Fisher,
Halperin and Morf \cite{Fisher} showed that the lowest energy defect states
correspond to interstitial defects, i.e. an extra electron sits on an interstitial
site in the otherwise perfect crystal. Single vacancies have a higher energy.
Following Fisher et al., there are two types of interstitial site, `centered' and
`edge' interstitials, and these are by far the most predominant type of defect at
finite temperatures. In our model, these defects can be created by introducing an
$(N+1)$'th electron which forms the defect. There are several reasons why this is
reasonable. First, the alternative scheme of allowing one of the $N$ existing
electrons to form the defect would create an interstitial {\em plus} vacancy;
following Fisher et al. the total energy of such a defect is approximately three
times larger than a single interstitial. Second, creation of such an
interstitial-vacancy pair would involve a transformation of both $\theta$ {\em
and} $\alpha$ coordinates within the $N$-electron $\Omega$-space. Third, the
definition of the $\theta_{[j]}$ variables (see Eq. (9)) is independent of the
coordinates of electron $N+1$. Hence the $N$-electron $\Omega'$ coordinate system
is essentially unchanged by the presence of the extra electron. The
hyperangular function $G(\Omega')$ for the
$N$ electron system can therefore be compared directly to the corresponding
hyperangular function for the $N+1$ electron system when projected onto the
$N$-electron $\Omega'$ space. We wish to consider the effect of this defect on
ring $m$. Following Fisher et al. \cite{Fisher} the  distortion of the crystal will
be well-localized around the defect. In terms of the hyperangular coordinates, part
of the local crystal distortion will be subsumed in the coordinate $r$ and 
the effect on the hyperangles $\alpha$ and $\theta$ of the electrons in ring $m$
will be relatively small unless the defect lies in ring $m$. Assume the defect
lies in ring $m=m_d$. The antisymmetry condition obtained above for the perfect
crystal ($J_{WX}$) will still be approximately valid for all rings with $m\neq
m_d$. In ring $m_d$, there are now an {\em odd} number of electrons $N_m+1$.  The
rotation  $\theta_j\rightarrow\theta_j+\frac{2\pi}{N_m+1}$ now corresponds to 
$\theta'\rightarrow\theta'+ \Delta\theta'$ where
$\Delta\theta'=(N_m+1)\frac{2\pi}{(N_m+1)N}= \frac{2\pi}{N}$ (N.B. we now have
an $(N+1)$ electron system). Because of the odd-member ring, rotation corresponds
to an even number of  $i\leftrightarrow j$ interchanges.  The hyperangular
function for the $N+1$ electron system of crystal plus
defect, when projected onto the original $N$-electron $\Omega'$-space, is
essentially unchanged -- the original $N$ electrons are only slightly distorted
by the presence of the defect. Hence the overall phase change
$e^{iJ\Delta\theta'}$ must equal $e^{i2\pi n'}$ where $n'$ is any integer.
Denoting the $J$ value as $J_{WXD}$ we therefore obtain the condition
\begin{equation} J_{WXD}=Nn' \ . \end{equation}  Again, this criterion for
$J_{WXD}$ is {\em independent} of $m$ and hence holds for a single defect located
in any ring $m$. Also,  $J_{WXD}$ must be an integer. These two
criteria, taken together, therefore guarantee that the $N$-electron wavefunction 
has the correct permutational symmetry under the subset of all permutations of $N$
electrons which  correspond to ring rotations, both for the perfect crystal (WX)
{\em and} the crystal plus defect (WXD). For the perfect crystal, we can consider
$N$ to be odd since each ring contains an even number of electrons, plus there is
one electron at the center. Combining the two conditions for $J_{WX}$ and
$J_{WXD}$ we hence see that the WX and WXD have the following common $J$ values:
\begin{equation} J_m=\frac{1}{2}N(N-1)(2n+1) \end{equation} where $n$ is any
integer. Converting these $J_m$-values into filling factors using the formula
$\nu=\frac{N(N-1)}{2J_m}$, which is valid for large $N$, yields \begin{equation}
\nu=\frac{1}{2n+1}\ . \end{equation} This coincides with the principal series of
FQHE fractions, i.e. $\frac{1}{3}$ and  $\frac{1}{5}$. The value $\nu=1$ will be
discussed below. As an illustration we consider the case of $N=201$ electrons. The
allowed $J_{WX}$ values are $100\times 1$, $100\times 3$, $100\times 5$ etc. while
the allowed  $J_{WXD}$ values are $201\times 1$, $201\times 2$, $201\times 3$ etc.
It is clear that common $J$ values are given by $J_m=100\times 201\times 1$,
$100\times 201\times 3$ etc. and hence $\nu=\frac{1}{3},\frac{1}{5}$.  

\vskip 0.2in

{\bf III.2 Spin-unpolarized system}

We have so far generated the $J_m$ values for a system of spin-polarized
particles. Next we consider the opposite limit of a spin-unpolarized system,
i.e. $N_{+}=N_{-}$ where $N=N_{+}+N_{-}$ and $N$
must therefore be an even number. The arguments will be more
approximate in this case, but we believe will still contain the essential physics.
Consider a `typical' ring as before. Let this ring, $m$, contain $N_m$ electrons
where
$N_m>>1$: the ring will typically have $\frac{N_m}{2}$ up-spins and $\frac{N_m}{2}$
down-spins.  Due to the Pauli principal keeping like spins apart, we will assume
that on the average the ordering corresponds to the alternating sequence
up-spin--down-spin repeated around the ring (See Fig. 6(a)). Rotation of the ring
to a topologically identical configuration now involves a rotation of all the
electrons in the ring by an angle $\frac{2\pi}{(N_m/2)}$, i.e. we have to rotate
through twice
$\frac{2\pi}{N_m}$. The rotation
$\theta_j\rightarrow\theta_j+\frac{2\pi}{(N_m/2)}$ corresponds to 
$\theta'\rightarrow\theta'+ \Delta\theta'$ where
$\Delta\theta'=N_m\frac{2\pi}{(N_m/2)(N-1)}= \frac{4\pi}{(N-1)}$. Since $N_m$ is
an even number, $\frac{N_m}{2}$ can either be odd or even. Rotation of the $m$'th
ring by $\frac{4\pi}{N_m}$ therefore corresponds to either an even or odd number
of interchanges  $i\leftrightarrow j$ for {\em both} spin-up and spin-down
electrons. Hence the total number of interchanges of {\em like} spins is always
even. The overall phase change must therefore equal $e^{i2\pi n}$ where $n$ is any
integer. Denoting the $J$ value as $J_{WX}$ we therefore obtain the condition
\begin{equation} J_{WX}=\frac{1}{2}(N-1) n \ . \end{equation}  Again this
criterion for $J_{WX}$ is independent of $m$ and hence holds for all rings $m$. 
Now consider the Wigner crystal plus defect (WXD) with the defect in  ring $m$.
The defect corresponds to an extra electron which can either be spin-up or spin-down. There
are now an {\em odd} number of electrons $N_m+1$.  We now have, for large $N_m$, that
$\Delta\theta'\approx 2(N_m+1)\frac{2\pi}{(N_m+1)N}= \frac{4\pi}{N}$ for the $N+1$ electron
system. Because of the odd-member ring, rotation now corresponds to an overall {\em odd}
number of  $i\leftrightarrow j$ interchanges. This is because either the spin-up
interchanges are odd while the spin-down are even, or vice versa. Hence the
overall phase change $e^{iJ\Delta\theta'}$ must equal $e^{i\pi (2n'+1)}$ where $n'$
is any integer. Denoting the $J$ value as $J_{WXD}$ we therefore obtain the
condition \begin{equation} J_{WXD}=\frac{1}{2}N (n'+\frac{1}{2}) \ .
\end{equation}  Again, this criterion for $J_{WXD}$ is independent of $m$. These
two criteria, taken together, therefore guarantee that the spin-unpolarized
$N$-electron wavefunction  has the correct permutational symmetry under the subset
of all permutations of $N$ electrons which  correspond to ring rotations, both for
the perfect crystal (WX) {\em and} the crystal plus defect (WXD). Combining the
two conditions for integer values of $J_{WX}$ and $J_{WXD}$ we hence see that the
WX and WXD have the following common $J$ values: \begin{equation}
J_m=\frac{1}{4}N(N-1)(2n+1) \end{equation} where $n$ is any integer. Converting
these $J_m$ values into filling factors yields \begin{equation}
\nu=\frac{2}{2n+1}\ . \end{equation} This coincides with the second series of FQHE
fractions, i.e. $\frac{2}{3}$, $\frac{2}{5}$, $\frac{2}{7}$ etc. and hence
suggests that the ground states at these fractions will be spin-unpolarized in
the absence of Zeeman energy, in agreement with earlier finite-size numerical
calculations (see p. 63 of Ref. 3). Interestingly the series also reproduces the
IQHE value $\nu=2$. As an illustration we consider the case of $N=200$ electrons.
The allowed $J_{WX}$ values are $199\times 1$, $199\times 2$, $199\times 3$ etc.
while the allowed  $J_{WXD}$ values are $50\times 1$, $50\times 3$, $50\times 5$
etc. Common $J$ values are given by $J_m=199\times 50\times 1$, $199\times
50\times 3$ etc. and hence $\nu=\frac{2}{3},\frac{2}{5}$ etc.  
   
\vskip 0.2in

{\bf III.3 Partially spin-polarized system}

Consider a partially spin-polarized system. First we will take
i.e. $N_{+}=3N_{-}$ where $N=N_{+}+N_{-}$ and $N$
is again even. Consider a `typical'
ring as before. Let this ring, $m$, contain $N_m$ electrons where $N_m>>1$: the
ring will typically have $\frac{3N_m}{4}$ up-spins and $\frac{N_m}{4}$ down-spins. 
Due to the Pauli principal keeping like spins apart, we will now assume that on the
average the ordering corresponds to the sequence
up-spin--up-spin--up-spin--down-spin repeated around the ring (see Fig. 6(c)).
Rotation of the ring to a topologically identical configuration now involves a
rotation of all the electrons in the ring by an angle $\frac{2\pi}{(N_m/4)}$, i.e.
we have to rotate through four times $\frac{2\pi}{N_m}$. The rotation
$\theta_j\rightarrow\theta_j+\frac{2\pi}{(N_m/4)}$ corresponds to 
$\theta'\rightarrow\theta'+ \Delta\theta'$ where
$\Delta\theta'=N_m\frac{2\pi}{(N_m/4)(N-1)}= \frac{8\pi}{(N-1)}$. Since $N_m$ is
an even number, $\frac{3N_m}{4}$ and $\frac{N_m}{4}$  are either both odd or both
even. Rotation of the $m$'th ring by $\frac{8\pi}{N_m}$ therefore corresponds to
either an even or odd number of interchanges  $i\leftrightarrow j$ for {\em both}
spin-up and spin-down electrons. Hence the total number of interchanges of {\em
like} spins is always even. The overall phase change must therefore equal
$e^{i2\pi n}$ where $n$ is any integer. Denoting the $J$ value as $J_{WX}$ we
therefore obtain the condition \begin{equation} J_{WX}=\frac{1}{4}(N-1) n \ .
\end{equation}  Again this criterion for $J_{WX}$ is independent of $m$ and hence
holds for all rings $m$. Now consider the Wigner crystal plus defect (WXD) with
the defect in  ring $m$. The defect can either be spin-up or spin-down. There are
now an {\em odd} number of electrons $N_m+1$.  We now have, for large $N_m$, that
$\Delta\theta'\approx 4(N_m+1)\frac{2\pi}{(N_m+1)N}= \frac{8\pi}{N}$ for the $N+1$
electron system. Because of the odd-member ring, rotation now corresponds to an
overall {\em odd} number of  $i\leftrightarrow j$ interchanges. This is because
either the spin-up interchanges are odd while the spin-down are even, or vice
versa. Hence the overall phase change $e^{iJ\Delta\theta'}$ must equal $e^{i\pi
(2n'+1)}$ where $n'$ is any integer. Denoting the $J$ value as $J_{WXD}$ we
therefore obtain the condition \begin{equation} J_{WXD}=\frac{1}{4}N
(n'+\frac{1}{2}) \ . \end{equation}  Again, this criterion for $J_{WXD}$ is
independent of $m$. These two criteria, taken together, therefore guarantee that
the partially spin-polarized $N$-electron wavefunction  has the correct
permutational symmetry under the subset of all permutations of $N$ electrons
which  correspond to ring rotations, both for the perfect crystal (WX) {\em and}
the crystal plus defect (WXD). Combining the two conditions for integer values of
$J_{WX}$ and $J_{WXD}$ we hence see that the WX and WXD have the following common
$J$ values: \begin{equation} J_m=\frac{1}{8}N(N-1)(2n+1) \end{equation} where $n$
is any integer. Converting these $J_m$-values into filling factors yields
\begin{equation} \nu=\frac{4}{2n+1}\ . \end{equation} This coincides with the
fourth series of FQHE fractions, i.e. $\frac{4}{5}$, $\frac{4}{7}$ etc.,
hence the theory suggests that the corresponding ground states at these fractions
will be partially spin-polarized in the ratio of spin-up to spin-down of $3:1$
in the absence of Zeeman energy. Again this is in agreement with finite-size
numerical calculations (see p. 63 of Ref. 3). As an illustration
we again consider the case of $N=200$ electrons (N.B. we have $4$ as a factor
since the ratio of spin-up to spin-down is $3:1$). The allowed $J_{WX}$ values are
$199\times 1$, $199\times 2$, $199\times 3$ etc. while the allowed  $J_{WXD}$
values are $25\times 1$, $25\times 3$, $25\times 5$ etc. Common $J$ values are
given by $J_m=199\times 25\times 1$, $199\times 25\times 3$ etc. and hence
$\nu=\frac{4}{5},\frac{4}{7}$ etc.  

Next we take
i.e. $N_{+}=2N_{-}$ where $N=N_{+}+N_{-}$ and $N$
is again even. Consider a `typical'
ring as before. Let this ring, $m$, contain $N_m$ electrons where $N_m>>1$: the
ring will typically have $\frac{2N_m}{3}$ up-spins and $\frac{N_m}{3}$ down-spins. 
Due to the Pauli principal keeping like spins apart, we will assume that on the
average the sequence corresponds to up-spin--up-spin--down-spin repeated
around the ring (see Fig. 6(b)). Rotation of the ring to a topologically identical
configuration now involves a rotation of all the electrons in the ring by an angle
$\frac{2\pi}{(N_m/3)}$, i.e. we have to rotate through three times
$\frac{2\pi}{N_m}$. The rotation
$\theta_j\rightarrow\theta_j+\frac{2\pi}{(N_m/3)}$ corresponds to 
$\theta'\rightarrow\theta'+ \Delta\theta'$ where
$\Delta\theta'=N_m\frac{2\pi}{(N_m/3)(N-1)}= \frac{6\pi}{(N-1)}$. Since $N_m$ is
an even number, $\frac{2N_m}{3}$ and $\frac{N_m}{3}$  are either both odd or both
even. Rotation of the $m$'th ring by $\frac{6\pi}{N_m}$ therefore corresponds to
either an even or odd number of interchanges  $i\leftrightarrow j$ for {\em both}
spin-up and spin-down electrons. Hence the total number of interchanges of {\em
like} spins is always even. The overall phase change must therefore equal
$e^{i2\pi n}$ where $n$ is any integer. Denoting the $J$ value as $J_{WX}$ we
therefore obtain the condition \begin{equation} J_{WX}=\frac{1}{3}(N-1) n \ .
\end{equation}  Again this criterion for $J_{WX}$ is independent of $m$ and hence
holds for all rings $m$. Now consider the Wigner crystal plus defect (WXD) with
the defect in  ring $m$. The defect can either be spin-up or spin-down. There are
now an {\em odd} number of electrons $N_m+1$.  We now have, for large $N_m$, that
$\Delta\theta'\approx 3(N_m+1)\frac{2\pi}{(N_m+1)N}= \frac{6\pi}{N}$ for the $N+1$
electron system. Because of the odd-member ring, rotation now corresponds to an
overall {\em odd} number of  $i\leftrightarrow j$ interchanges. This is because
either the spin-up interchanges are odd while the spin-down are even, or vice
versa. Hence the overall phase change $e^{iJ\Delta\theta'}$ must equal $e^{i\pi
(2n'+1)}$ where $n'$ is any integer. Denoting the $J$ value as $J_{WXD}$ we
therefore obtain the condition \begin{equation} J_{WXD}=\frac{1}{3}N
(n'+\frac{1}{2}) \ . \end{equation}  Again, this criterion for $J_{WXD}$ in
independent of $m$. These two criteria, taken together, therefore guarantee that
the partially spin-polarized $N$-electron wavefunction  has the correct
permutational symmetry under the subset of all permutations of $N$ electrons
which  correspond to ring rotations, both for the perfect crystal (WX) {\em and}
the crystal plus defect (WXD). Combining the two conditions for integer
values of $J_{WX}$ and $J_{WXD}$ we hence see that the WX and WXD have the
following common $J$ values: \begin{equation} J_m=\frac{1}{6}N(N-1)(2n+1)
\end{equation} where $n$ is any integer. Converting these $J_m$ values into
filling factors yields \begin{equation} \nu=\frac{3}{2n+1}\ . \end{equation} This
coincides with the third series of FQHE fractions, i.e. $\frac{3}{5}$,
$\frac{3}{7}$ etc. and predicts the corresponding ground states to be partially
spin-polarized in the ratio of spin-up to spin-down of $2:1$ (or vice versa) in
the absence of Zeeman energy.  We 
note that there is an alternative system that also yields the filling factor
series $\nu=\frac{3}{2n+1}$. In particular, this fraction emerges from
considering a fully spin-polarized system but now considering rotation of two
rings simultaneously. These two states with $N_+:N_-=2:1$ and $N_-=0$
respectively, probably compete to become the ground-state depending on the value
of the Zeeman energy (and hence magnetic field). Interestingly, finite-size studies
have shown that the $\nu=\frac{3}{5}$ state is indeed partially polarized for
$B<15$T in the ratio $N_+:N_-=2:1$ but fully polarized for $B>15$T (see
p.160 of Ref. 3). 

The above arguments can be extended straightforwardly to 
consider $N_{+}=(p-1)N_{-}$ where $N=N_{+}+N_{-}$ and $p$ is any integer. In this
case, the corresponding filling fraction becomes $\nu=\frac{p}{2n+1}$. We
now focus on the filling factor $\nu=1$ (IQHE). We have already shown that
this state emerges from the two following series:  $\nu=\frac{1}{2n+1}$
for $n=0$ in a fully spin-polarized system, or  $\nu=\frac{3}{2n+1}$ for $n=1$ in
a partially spin-polarized system. In fact for a given spin polarization
$N_{+}=(p-1)N_{-}$ with $p$ odd, the factor $\nu=1$ will always arise, i.e. by
choosing  $n=\frac{p-1}{2}$. It is reasonable to expect these states to compete to
become the ground-state. 
This is consistent with recent findings that a gap exists at $\nu=1$ (IQHE) even
in the {\em absence} of Zeeman splitting. The possible coexistence of a
manifold of partially spin-polarized states is also
consistent with the idea of macroscopic spin-textures near $\nu=1$;
in particular, taking a linear combination of partially spin-polarized states
enables the construction of localized `wave-packets' of spin -- we conjecture that
the resulting spin-textures may be related to skyrmions.

 We note that the above
filling factors emerged from requiring that the WX and WXD had a common angular
momentum value. The defect was considered to be an interstitial electron.
It turns out that, as far as the common $J_m$ values are concerned, we could
also have considered the defect to be a vacancy. The product $(N-1)(N-2)$
would have appeared throughout this section instead of $N(N-1)$; in the
large $N$ limit, both products yield $\sim N^2$ and hence the same filling
factors. Hence at $J=J_m$, the WX can coexist with a WXD where the defect is either
an interstitial electron or a vacancy. As noted earlier, however, such vacancies do
have a higher energy and hence are less likely to occur at finite
temperatures.

\vskip 0.2in

{\bf IV. ANALYTIC CALCULATION OF FQHE ENERGY GAPS}

In this section we will give an analytic calculation for the energy gaps
associated with the FQHE and IQHE states, i.e. for $J=J_m$. The calculation is
approximate since it relies on the various approximations made in Sec. II.3.
However our goal is to find whether the gaps predicted by our model are in fact
consistent with the observed FQHE gaps and also to identify trends in the energy
gaps with filling fraction, magnetic field etc. 

In Sec. II.3 we obtained an approximate expression for the relative energy
$E_{\rm rel}$ (see Eq. (27)). This energy depends on $E_\alpha$ and $E_\theta$.
In Sec. III we argued that the important criterion characterising the
magic-number $J$-values was that the crystal (WX) and defect (WXD) can both
have the same $J$ value given by $J=J_m$; this leads to a large delocalization
energy due to increased WX-WXD tunnelling in $\Omega'$-space. In the language
of single-particle tight-binding theory, the resulting energy gap between states
with $J=J_m$ and $J\neq J_m$ arises from the
hybridization of the $G(\Omega')$ solutions peaked around, for example,
$\Omega'_0$ and $\Omega'_{0;a}$ at $J=J_m$. This hybridization hence yields a low
$E_{\rm rel}$ because of the corresponding delocalization of $G(\Omega')$ at
$J=J_m$, i.e. a reduction in zero-point energy. Here we will obtain an analytic
expression for this energy using a simple model for the effect of
delocalization and show that the resulting gaps are consistent with 
experimental findings. 

As pointed out earlier, a state of a given negative $J$ will have an energy minimum
at a finite magnetic field $\omega_c$. As $\omega_c$ increases, the value of $J$ at
which the energy $E_{rel}(J)$ has a minimum will increase. We will calculate the
energy difference between a state with $J=J_m$ and competing low-energy states with
$J$ given by $J_{m\pm}=J_m\pm \delta$ at  a given $\omega_c$, where $\delta<<J_m$.
The lowest energy state with $J=J_m$ is
given by Eq. (27) with $n=0$:  \begin{equation} E_{\rm
rel}(J_m)=\hbar\omega_0(B)\big[
\big([N-2]^2+J_m^2+\frac{2m^*\beta}{\hbar^2}V(\Omega'_0) +
{\frac{2m^*}{\hbar^2}}[E_\theta(J_m)+E_\alpha(J_m)] \big)^{\frac{1}{2}} +1\big]\\ 
-J_m\frac{\hbar\omega_c}{2}\  \end{equation}  while that with $J=J_{m\pm}$ is given
by:  $$E_{\rm rel}(J_{m\pm})=\hbar\omega_0(B)\big[
\big([N-2]^2+[J_{m}\pm\delta]^2+$$
\begin{equation} \frac{2m^*\beta}{\hbar^2}V(\Omega'_0) +
{\frac{2m^*}{\hbar^2}}[E_\theta(J_{m\pm})+E_\alpha(J_{m\pm})] \big)^{\frac{1}{2}}
+1\big]\\  -[J_{m}\pm\delta]\frac{\hbar\omega_c}{2}\ \ . \end{equation}  As
discussed in Sec. II.1, the angular momentum is negative for low-lying energy
states, hence we have included the minus sign directly into these expressions, i.e.
$J_m$ and $J_{m\pm}$ are positive numbers.  Consider the state with $J=J_m$. This
state can coexist as both a crystal (WX) and a crystal plus defect (WXD), in
contrast to the state $J_{m\pm}$. The transition WX$\rightarrow$WXD will involve a 
distortion of the $\Omega'$-coordinates.  This distortion in $\Omega'$-space
corresponds to a spreading or `delocalization' of the function $G(\Omega')$ along
these directions. There is therefore a reduction in `localization' energy going
from a state with $J\neq J_m$ (i.e. WX and WXD cannot coexist) to a state with
$J=J_{m}$ (i.e. WX and WXD can coexist).  We will hence write:\begin{equation}
E_\theta(J_{m})+E_\alpha(J_{m})+\beta V(\Omega'_0)= E_0 \end{equation} while
\begin{equation} E_\theta(J_{m\pm})+E_\alpha(J_{m\pm})+\beta V(\Omega'_0)
=E_0 + {\tilde\Delta}
\end{equation} where ${\tilde\Delta}$ represents the increased localization energy
of state $J_{m\pm}$ as compared to $J_m$. Note that the potential energy
minimum $\beta V(\Omega'_0)$ is a constant term throughout. In the Appendix we 
show that typically ${\tilde\Delta}\sim N^2$ while $E_0\sim N^3$. Hence in the
limit of $N>>1$, we have $E_0>>{\tilde\Delta}$. We also recall from Sec. III that
$J_m\sim N^2$.  It follows by expanding out  Eqs. (45) and (46) in the limit
$N>>1$ that 
\begin{equation} \Delta E_\nu\equiv
E_{rel}(J_{m\pm})-E_{rel}(J_m)\approx 
\hbar\omega_0(B)
\bigg[\frac{ \frac{m^*{\tilde\Delta}}{\hbar^2}\pm
J_m\delta}{[J_m^2+\frac{2m^*E_0}{\hbar^2}]^{\frac{1}{2}}}
\bigg]\mp\delta\frac{\hbar\omega_c}{2}\ .  \end{equation} 
We are interested in the large $N$ limit since our goal is to calculate the FQHE
gaps: hence we will choose the confinement
$\omega_0<<\omega_c$ which yields $\omega_0(B)\sim
\frac{\omega_{c}}{2}$. Let us first consider the filling factor $\nu=\frac{1}{3}$
hence
$J_m=3\frac{N(N-1)}{2}$. Substituting into Eq. (49) we obtain the
approximate expression for the energy gap  \begin{equation}
\Delta
E_{\frac{1}{3}}\sim\frac{1}{3}\frac{m^*{\tilde\Delta}}
{N^2\hbar^2}\hbar\omega_{c}\ .
\end{equation} 
There are several points to note about this expression for the energy gap at
$\nu=\frac{1}{3}$. 

\noindent (i) Given that ${\tilde\Delta}\sim N^2$ for large $N$, the expression
is {\em independent} of the electron number $N$. It is also independent of
the strength of the parabolic potential $\omega_0$. For a given
$\omega_{c}$ corresponding to the filling factor $\nu=\frac{1}{3}$, we can
therefore take the thermodynamic limit $N\rightarrow\infty$ and yet still maintain
a {\em fixed} average electron density 
by choosing appropriately small values of $\omega_0$. Our expression
for the energy gap at filling factor $\nu=\frac{1}{3}$ which was derived in terms
of $J$ values for a fixed-$N$ system therefore also holds for an infinite
two-dimensional electron gas of fixed density. 

\noindent (ii) The expression for the
energy gap does not exhibit a direct dependence on the value of the
electron-electron interaction $\beta$. We do emphasize, however, that throughout
most of this paper we have {\em assumed} that $\beta$ is large enough to be able
to neglect tunnelling between SEMs. Hence we can only conclude that 
the absolute value of $\beta$ does not directly
affect the energy gap $\Delta E$ for
sufficiently large $\beta$. This is consistent with experimental findings
that the gap can be remarkably sample-independent \cite{Chakraborty}. Below we
will mention how a weak dependence on $\beta$ will arise if one
considers smaller values of $\beta$. 

\noindent (iii) The energy gap appears to be approximately {\em linear} in the
magnetic field. Most previous theoretical studies conclude that the dependence
resembles
$B^\frac{1}{2}$. As we will show below, the linear dependence is in
reasonable agreement with experimental data, particularly at lower fields. However,
we will later  discuss how a weaker, non-linear dependence, i.e. $B^x$ where $x<1$,
can eventually arise at larger $B$ in our model.

\noindent (iv) The energy gap does not depend on $\delta$ to first order in
$(\frac{\delta}{J_m})$. This independence of $\delta$ is important since it implies
that the energy gap exists between state $J_m$ and {\em all} other states with $J$
in the vicinity of
$J_m$. We know from the discussion at the end of Sec. II.1 that, over a given range
of magnetic field, the  states  which compete to become the ground state will
be those of similar
$J$. Hence we expect the energy gap arising to exist over a small but finite range
of magnetic field, as observed experimentally. 

\noindent (v) The expression for the gap $\Delta E$ can be made applicable to
situations where the lowest-energy excitation involves spin-flips, by adding
$\Delta E_{spin}$ where
$\Delta E_{spin}$ is the difference in total spin energy between the excited state
and the
$\nu=\frac{1}{3}$ spin-polarized ground state. However, we are
interested in the lowest-energy excitations, and hence will take $\Delta
E_{spin}=0$ since $\Delta E_{spin}\geq 0$ for the fully spin-polarized initial
state at $\nu=\frac{1}{3}$. For other fractions, the term
${\tilde\Delta}$ will have an indirect dependence on the spin configuration
since, for a mixed spin system, the $\Omega'$ space is really coupled to the spinor
space by the antisymmetry condition, hence $G(\Omega')$ is actually a
two-dimensional vector. A detailed discussion of spin-reversed excited states will
be given elsewhere.

\noindent (vi) Given the discussion in (v), together with the fact that 
${\tilde\Delta}$ has a weak but finite dependence on $J_m$ (see later), we can
conclude the value of ${\tilde\Delta}$ will generally be different for different
fractions. Gaps at other fractions in the
$\nu=\frac{p}{3}$ series are discussed below.

We will now attempt to derive an approximate analytic expression for
${\tilde\Delta}$.  Consider the
$N$-electron system at $J=J_m$. As discussed in Sec. III, the system can exist as
both a crystal (WX) and a crystal plus defect (WXD), in contrast to the state
$J_{m\pm}$. We will argue in what follows that the transition WX$\rightarrow$WXD
involves a significant fractional distortion of $\theta$-coordinates.  This
distortion along the $\{\theta_{[j]}\}$-axes in $\Omega'$-space corresponds to a
spreading or `delocalization' of the function $G(\Omega')$ along these directions,
thereby giving rise to a finite ${\tilde\Delta}$. 
In Secs. II and III, we argued that the low-energy SIMs near a given SEM consisted
of a  single electron defect placed at an  interstitial site within the hexagonal
crystal. As noted at the end of Sec. III the defect could also be a vacancy,
although the corresponding SIM would have a higher energy. 
Consider a defect 
placed in ring $m$ which contains $N_m$ particles with $N_m>>1$, and let the
defect be sited between $j$ and $j+1$ (c.f. Fig. 5). 
Classically, the system moves to a nearby SIM in $\Omega'$-space (c.f. Fig. 2). 
In particular the defect will cause a distortion of the coordinates of particle $j$.
In order to calculate the maximum possible distortion (and hence delocalization
available as a result of the hybridization between the SEM and SIM) we will consider
the particular SIM in which only particle $j$ moves to accomodate the defect. In
principle, both the $\alpha_j$ and $\theta_{[j]}$ coordinates will be modified,
thereby `sharing' the effect of the distortion. However, with the defect placed
between $j$ and $j+1$ in ring $m$, the distortion of particle $j$ will mainly be
along the $\theta_{[j]}$ direction.  The idea that the most important effect of
the defect is  the distortion of the $\theta$-coordinates, is consistent with the
following considerations. Consider any particle $j'$ nearby to the defect with
coordinates $\theta_{[j']}$ and $\alpha_{j'}$. Let the nearby defect cause a
distortion of $a$ in all directions. Hence the new coordinates of the particle $j'$
are approximately $\alpha_{j'}+a$ and $\theta_{[j']}+\frac{a}{\alpha_{j'}}$. The
relative distortion caused by the nearby defect is hence
$\frac{\Delta\alpha_{j'}}{\alpha_{j'}}\sim \frac{a}{\alpha_{j'}}$ while
$\frac{\Delta\theta_{[j']}}{\theta_{[j']}}\sim \frac{a}{\alpha_{j'}\theta_{[j']}}$.
For $N>>j'$, $\alpha_{j'}$ changes slowly with $j'$ and is unchanged if $j'$ lies in
ring $m$. However $\theta_{[j']}$ ranges from $0$ to $2\pi$ within ring $m$ and hence
the fractional change $\Delta\theta_{[j']}$ can be significant.  

We will therefore consider
${\tilde\Delta}$ as arising as a result of the difference in $E_\theta$ for $J_m$ as
compared to $J_{m\pm}$. The loss of
`localization' energy of the state $J_m$ as compared to $J_{m\pm}$  can be
significant along the $\{\theta_{[j]}\}$-directions, i.e. 
$E_\theta$ can differ appreciably depending on whether the function $G(\Omega')$ is
localized (i.e. $J=J_{m\pm}$) or not ($J=J_m$). Consider Eq. (26) for the `energy'
$e_j$ associated with a particle $j$ in ring $m$ where $J\neq J_m$ (i.e. WX and WXD
cannot coexist). The equation resembles a one-dimensional Schrodinger equation in
$\theta_{[j]}$ for a mass $\alpha_{j0}$ moving in a potential
$w_j(\theta_{[j]}-\theta_{[j0]})$. The function $w_j$ will have a minimum at
$\theta_{[j]}=\theta_{[j0]}$, whereas there are maxima at 
$\theta_{[j]}=\theta_{[j0]}\pm\frac{2\pi}{N_m}$ as particle $j$ approaches particle
$j\pm1$. For $N=3$, recall Fig. 3 where moving along the $y$-axis ($\theta$-axis) at
fixed $\alpha=\alpha_0$ (i.e. $x=0$) produced a minimum at $y=0$ and maxima at
$y=\pm\frac{\pi}{2}$. We can approximate $e_j$ using a simple one-dimensional
particle-in-a-box model: we will assume that  $w_j$ is a flat-bottomed potential
with infinite walls at $\theta_{[j]}=\theta_{[j0]}\pm \frac{2\pi}{N_m}$. The width
of the box is therefore given by $a_j=\frac{4\pi}{N_m}$. The energy $e^{(J)}_j$ is 
hence approximately given by \begin{equation} e^{(J)}_j\sim \frac{\hbar^2 \pi^2
N_m^2}{2m^*\alpha^2_{j0} [4\pi]^2}\ .  \end{equation} Note that the width of the
wavefunction $g_j$, and hence the localization of $G(\Omega')$ along $\theta_{[j]}$,
is characterized by $a_j$. This expression (Eq. (51)) for $e^{(J)}_j$ implicitly
assumes that the electron-electron interaction $\beta$ is large; for smaller values
of $\beta$, the particle-in-a-box energy should pick up a weak dependence on
$\beta$. Throughout this paper, however, we will use the approximate analytic form
given in Eq. (51).  At $J=J_m$ there is distortion of $G(\Omega')$ since WX and WXD
can coexist. The defect can occupy any interstitial site in the crystal; each defect
position produces a distinct SIM. There exists a SIM (e.g. $\Omega'_{0;j}$) in which
the defect is placed next to particle $j$, say between particles $j$ and
$j+1$ as used earlier. We can see that there is one such SIM
associated with each  $\theta_{[j]}$ coordinate. If, as discussed above, we consider
the main distortion as occuring on the $\theta_{[j]}$ coordinate of particle $j$,
then the associated SIM lies on the $\theta_{[j]}$ axis. In this case the
coexistence of the SEM $\Omega'_0$ and the SIM $\Omega'_{0;j}$ causes an increase in
the effective box-width,  $a_j\rightarrow a_j+\delta a_j$. The energy along the
$\theta_{[j]}$ direction is now given by   \begin{equation}e^{(Jm)}_j\sim e^{(J)}_j
+ \frac{\partial e^{(J)}_j}{\partial a_j}\delta a_j \sim e^{(J)}_j[1-2\frac{\delta
a_j}{a_j}]\ . 
\end{equation}   Although there are many such SIMs associated with each SEM, we know
that the defects have a low density at the temperatures of interest. The system also
requires a large time to tunnel between these SIMs since each SIM describes a
different defect position in the crystal: diffusion of the defect between sites will
be slow at low temperatures. It is reasonable therefore to suppose that each SEM
hybridizes with just one of these SIMs at any time. The average loss of localization
energy of state $J_m$ as a result of distortion due to a nearby SIM is therefore
obtained by averaging over all $(N-2)$ $\theta_{[j]}$ coordinates. Hence using Eq.
(52)\begin{equation} {\tilde\Delta} \sim \frac{1}{N-2}\sum_j 2\big[\frac{\delta
a_j}{a_j}\big] e^{(J)}_j \sim  2\overline{\big[\frac{{\delta a}}{a}\big]} {\bar e} 
\end{equation}  where $\bar x$ represents an average of the quantity $x$ over all
$(N-2)$ of the  $\theta_{[j]}$ coordinates.   

We could also try to obtain  expressions for $\Delta E$ at other fractions.
Consider $\nu=\frac{1}{5}$. Equation (50) now has the factor $\frac{1}{3}$ replaced
by $\frac{1}{5}$. Assuming as a crude approximation that the values of 
${\tilde\Delta}$ are the same, we obtain the result
that
$\Delta E_{\frac{1}{5}}:\Delta E_{\frac{1}{3}}\sim 0.6$ for samples at a given
magnetic field. The literature tends to put this ratio at about $0.3-0.4$
\cite{Boebinger,Robin}. 
One could also try to evaluate
$\Delta E$ for other fractions where the ground-state
$J_m$ and/or the excited states $J_{m\pm}$ are thought to have spin-reversed
electrons. Such a calculation needs a more careful estimation of
${\tilde\Delta}$, as discussed earlier. Here we will just provide a rough estimate
by considering, as before, excitations which do not change the total spin component
(i.e. $\Delta E_{spin}=0$). We will choose
$\nu=\frac{2}{3}$ in order to compare with the results for $\nu=\frac{1}{3}$.
There are two effects of considering $\nu=\frac{2}{3}$ instead of
$\nu=\frac{1}{3}$ when re-deriving the expressions obtained in this section.
The prefactor in Equation (50) increases by a factor of $2$, while $\frac{\delta
a_j}{a_j}$ decreases by a factor of $2$. This decrease in $\frac{\delta a_j}{a_j}$
arises as follows.  Recall that like spins in the
spin unpolarized case are separated by twice the angle of the spin-polarized
case. The `unit-cell' size in $\Omega'$-space is determined by the separation
between neighboring SEMs, and is therefore twice as large for the unpolarized
case. The effective box size must therefore also be twice as large, i.e.
the effective $a_j$ value is now
$\frac{8\pi}{N_m}$ instead of $\frac{4\pi}{N_m}$. 
Hence the final
result is that
$\Delta E_{\frac{2}{3}}\sim 
\Delta E_{\frac{1}{3}}$. Although we do not attach too much importance to this
result because of the complications of spin, it is interesting to note that
experimentally the gaps for $\nu=\frac{1}{3}$ and $\nu=\frac{2}{3}$ are found to be
similar \cite{Chakraborty,Boebinger,Robin} as will be shown below. 
We note that allowing for spin-flip excitations at $\nu=\frac{2}{3}$ may
reduce the overall gap at low $B$ since $\Delta E_{spin}$ can be negative if the
ground state is not fully spin-polarized. Hence the total gap may be negligible
for $\nu=\frac{2}{3}$ at low $B$. This feature is also seen experimentally. The
same argument concerning effective box size should also be approximately true for
the other partially spin-polarized fractions in the
$\frac{p}{3}$ series. Consider $\nu=\frac{p}{3}$. The prefactor in Equation (50)
increases by a factor $p$, while the effective box-size also increases by the
same factor. The net effect is that the expression for the gap is similar for all
fractions $\frac{p}{3}$ where $p=1,2,4,5$ etc. Hence the energy gaps for the
$\frac{p}{3}$ fractions measured across a range of samples should all fall on
approximately the same curve as a function of magnetic field.

Equation (53) presented an analytic expression for ${\tilde\Delta}$ which, to the
level of approximation employed, did not depend on the ground state $J_m$ value.
Examination of the more accurate versions of the hyperangular equation presented in
Sec. II.3, suggests that ${\tilde\Delta}$ should actually have a weak but finite
dependence on $J_m$ (recall, for example, Eq. (23) or (24)). In the $N=3$  study in
Sec. II.2, we found that increasing $J$ did indeed increase the localization  of
$G(\Omega')$ around the classical minima. For large $N$, as discussed in Sec. II.3,
the dependence on $J_m$  should be weaker; however the resulting hybridization
between a given SEM (i.e. WX) and nearest-neighbor SIMs (i.e. WXD) should also 
decrease
slightly as $J_m$ increases. 
Decreasing hybridization will reduce the value of ${\tilde\Delta}$ as $J_m$ 
increases. 
It is interesting to analyze the effect of this reduction in ${\tilde\Delta}$ for 
the separate situations of (a)
a given sample over a range of $\nu$ values, and
(b) different samples at a given fixed $\nu$. In case (a), the number of electrons
$N$ is fixed. As the value of $J_m$ increases, the value
of $\omega_c$ at which this $J_m$ value represents the ground state must
also increase. If ${\tilde\Delta}$ decreases as $J_m$  increases, then
${\tilde\Delta}$ must also decrease with increasing $\omega_c$.  This reduction in 
${\tilde\Delta}$ as $\omega_{c}$ increases, if sufficiently large, will make $\Delta
E\rightarrow 0$: we suggest that this may be related to the predicted formation of a
Wigner solid (i.e. gapless excitations) at very high magnetic fields.  In case (b),
fixed $\nu$ means that increasing $J_m$ requires an increase in $N$ (recall that
$J_m=\frac{N(N-1)}{2\nu}$).  If ${\tilde\Delta}$ decreases as $J_m$ 
increases, then ${\tilde\Delta}$ must also decrease as $N$ increases. However, a
fixed value of $\nu$ means that increasing $N$ requires an increase in
magnetic field. Hence ${\tilde\Delta}$ decreases as $\omega_c$ 
increases. 
As mentioned earlier, this will tend to weaken the linear
magnetic field-dependence of the theoretical gaps at fixed $\nu$ to $\omega_c^x$ with
$0<x<1$. Such a sub-linear dependence  is consistent with recent
experimental data at high fields \cite{Robin2}. In the estimates of the
gap discussed below, however, we take as a first approximation the form of
${\tilde\Delta}$ presented in Eq. (53). Consequently the calculated gaps for a
given $\nu$ always increase linearly with magnetic field.

We now proceed to discuss appropriate values of $\overline{\big[\frac{{\delta
a}}{a}\big]}$ and hence calculate the gaps. The precise value for
$\overline{\big[\frac{{\delta a}}{a}\big]}$ will depend on the details of the
crystal plus defect system (WXD). Here we will suggest reasonable lower and upper
estimates, and argue that the particular value to be used will depend on the degree
of disorder in the experimental samples. In particular, we will argue that the
lower estimate is appropriate for disordered (i.e. lower mobility) samples 
while the
upper estimate is appropriate for pure (i.e. higher mobility) samples. 

First
we consider the lower estimate for $\overline{\big[\frac{{\delta a}}{a}\big]}$. Fisher et al. found that the maximum
distortion for a vacancy defect was about
$12\%$, but that the value for the interstitial defect was `considerably larger'
\cite{Fisher}.
If the sample contains a significant impurity concentration, it is likely that
interstitial electrons will have difficulty in diffusing through the
$N$-electron system. The function $G(\Omega')$ will therefore have a restricted
delocalization for kinetic reasons. In the absence of interstitial defects, the
delocalization would be determined solely by the vacancies.  We will therefore take
the value of
$12\%$ as a lower estimated bound for 
$\overline{\big[\frac{{\delta a}}{a}\big]}$. From the
Appendix we have that
${\bar e}\sim \frac{27\hbar^2 N^2}{16m^*\pi^2}$ and hence ${\tilde\Delta}\sim
0.4\frac{\hbar^2 N^2}{m^*\pi^2}$. Substituting this into the expression for the
energy gap, we obtain $\Delta E_{\frac{1}{3}}\sim 0.014\hbar\omega_{c}$ meV and
hence $\Delta E_{\frac{1}{3}}\sim 0.16\hbar\omega_{c}$ Kelvin.   Given that
$1$meV$\equiv 1.728B[{\rm T}]$ (Tesla) for GaAs, we obtain $\Delta
E_{\frac{1}{3}}\sim 0.27B[{\rm T}]$ Kelvin. Hence at $B=20$T, $\Delta
E_{\frac{1}{3}}\sim 5.5$ Kelvin.  
Figure 7 compares this lower estimated bound of the energy gap at
$\nu=\frac{1}{3}$ (and hence $\nu=\frac{2}{3}$ as explained above)
with early experimental results obtained by Boebinger
et al. \cite{Boebinger,Chakraborty} over a range of relatively impure samples
(i.e. significant impurity concentration). The agreement is surprisingly good;
however we emphasize that our calculation is obviously fairly crude. Apart from
improving the expression for
$\Delta E_{\frac{1}{3}}$ given in Eq. (50), one could do a better job in
calculating the localization energy ${\tilde\Delta}$ from Eq. (53). Such
improvements would almost certainly render the calculation of energy gaps within
the present model numerical, although such calculations would be more
straighforward than the original alternative of an $N$-electron diagonalization.
Various numerical improvements will be presented in a future publication; the goal
of the present paper is to pursue a purely analytical theory.

Now we turn to an upper estimate for $\overline{\big[\frac{{\delta
a}}{a}\big]}$. We assume that the sample is pure and hence there is no kinetic
reason for ignoring interstitial defects. 
We first recall our physical picture of the topology of the interstitial defect,
stated earlier in this section. We consider a ring $m$
containing $N_m$ electrons, including particles $j-1$,
$j$ and
$j+1$ (c.f. Fig. 5). The angle between particle $j$ and $j+1$ is 
$\frac{2\pi}{N_m}$, and the angle between particle $j-1$ and $j+1$ is
$\frac{4\pi}{N_m}$. We let the defect lie in the ring between
particles
$j$ and $j+1$. We are looking for an upper estimate on the value of
$\overline{\big[\frac{{\delta a}}{a}\big]}$, hence we will assume that
the only particle which moves to accomodate the extra electron is particle $j$. As
before, the angle between 
$j-1$ and $j+1$ is still $\frac{4\pi}{N_m}$, but now there are {\em two} particles
($j$ and the defect) within this angular range. In an equilibrium state (i.e. a SIM)
the three angles between $j-1$ and $j$, $j$ and the defect, and the defect and
$j+1$, are all equal to $\frac{4\pi}{3N_m}$. The distortion of the effective box
size for particle $j$ will be determined by the angle between $j$ and $j+1$. Hence
an estimate of the average distortion $\overline{\big[\frac{{\delta
a}}{a}\big]}$ is $\frac{2\pi}{N_m}(\frac{4}{3}-1)$ divided by $\frac{2\pi}{N_m}$,
which gives
$\frac{1}{3}$. Our upper estimated value of $\overline{\big[\frac{{\delta
a}}{a}\big]}$ is therefore $0.33$. Figure 8 
compares both this upper bound and the lower bound obtained
earlier to experimental data obtained by the Oxford and $\rm{AT\& T}$
groups for a range of relatively pure, high mobility samples
\cite{Robin,Willett}.  The experimental values lie between the two bounds. This
consistency between the  present theoretical results and experiment
lends support to our interpretation of the effect of sample purity.

\vskip 0.2in

{\bf V. CONCLUSIONS}

A microscopic theory describing a confined
$N$-electron gas in two dimensions subject to an external magnetic field, was 
presented.  The
number of electrons $N$ and strength of the electron-electron interaction can be arbitrarily
large. For any value of the magnetic field $B$, the correlated $N$-electron states were
shown to be determined by the solution to a universal effective problem: this problem
resembles that of a ficticious particle moving in a multidimensional space, without a
magnetic field, occupied by potential minima corresponding to the classical $N$-electron
equilibrium configurations. 

A
possible connection with the 
Fractional (FQHE) and Integer  (IQHE) Quantum Hall effects was subsequently
proposed. In particular, it was shown that low-energy minima can arise in the large
$N$ limit at filling factors $\nu=\frac{p}{2n+1}$ where $p$ and $n$ are any
positive integers. The energy gaps calculated analytically at $\nu = \frac{p}{3}$
were found to be consistent with experimental data as a function of
magnetic field, over a range of samples. Various other known features of FQHE and
IQHE  states were found to emerge from the present theory. 

While it is obviously extremely difficult to calculate
many-particle energy gaps etc. accurately using an analytic approach, we hope that the
general qualitative trends and orders of magnitude provided by the model will be useful in
understanding the fascinating but complex field of highly-correlated $N$-electron systems.
We also hope that the model may begin to shed some light on the connection between the two
limits of few-electron correlated states in quantum dots, and the infinite two-dimensional
electron gas.

\vskip 0.2in

{\bf ACKNOWLEDGEMENTS}

This work was supported by EPSRC Grant No. GR/K 15619  (U.K.) and by COLCIENCIAS Project No.
1204-05-264-94 (Colombia). We thank Prof. Robin
Nicholas and Dr. V.N. Nicopoulos for useful comments toward the latter stages of this work,
and Dr. P.A. Maksym for an earlier discussion. N.F.J. also thanks Prof. P.M. Hui for
discussing his unpublished results on few-electron classical configurations during a
joint collaboration financed by the British Council under the UK-Hong Kong Joint Research
Scheme.

\newpage
\vskip 0.2in

{\bf APPENDIX}

First we derive an approximate expression for $\bar e$ and hence ${\tilde\Delta}$.
The exact hyperangular identity $\sum_{j=2}^N [\frac{X_j}{r}]^2=1$ implies $\sum
\alpha_j^2\sim 1$.  The moment of inertia $I$ of the $(N-1)$-particle system in
$\Omega$-space, treated as a rigid body, is therefore approximately just $m^*$.
For large $N$, the density of particles will be approximately uniform: the moment
of inertia for such a uniform disk is just $\frac{1}{2}(N-1)m^*R^2\sim 
\frac{1}{2}Nm^*R^2$,
where $R$ represents the disk radius and $(N-1)m^*$ is the total mass. Hence
$R^2\sim\frac{2}{N}$, the average density of particles is $\frac{(N-1)}{\pi
R^2}\sim \frac{N^2}{2\pi}$, and the average particle-particle spacing is $\sim
\frac{[2\pi]^\frac{1}{2}}{N}$. 
Now consider the sum over energies $e^{(J)}_j$ from Eq. (51):
\begin{equation}
\sum e^{(J)}_j\sim \sum \frac{\hbar^2 \pi^2 [N_m]^2}{2m^*\alpha^2_{j0}
[4\pi]^2}\ .
\end{equation}
The quantity $\alpha_{j0}$ for particle $j$ in ring $m$ is approximately 
$m$ times the average particle-particle separation: i.e. 
$\alpha_{j0}\sim m\frac{[2\pi]^\frac{1}{2}}{N}$.
Replacing the sum over $j$ by a sum over the rings $m$, and using the approximate
result that there are $6m$ particles in ring $m$, yields  \begin{equation}
\sum_j e^{(J)}_j\sim \sum_m [6m]^3\frac{\hbar^2 N^2}{64m^*\pi m^2}
\sim \frac{27\hbar^2 N^2}{8m^*\pi}\sum_1^{m_{max}} m
\end{equation}
where $m_{max}$ is the maximum ring number. $m_{max}$ is given approximately
by the disk radius $R$ divided by the particle-particle separation. Hence
$m^2_{max}\sim \frac{N}{\pi}$. For large $m_{max}$, $\sum_1^{m_{max}} m\sim
\frac{1}{2}m^2_{max}$. Hence 
\begin{equation}
\sum e^{(J)}_j\sim \frac{27\hbar^2 N^3}{16m^*\pi^2}\ .
\end{equation}
We require the average $e^{(J)}_j$ value, $\bar e$, with the average taken over all
$j$. There are $N-2$ such $j$ coordinates, hence for large $N$ we have ${\bar
e}\sim  \frac{27\hbar^2 N^2}{16m^*\pi^2}$ as claimed in Sec. IV. Given that 
\begin{equation} {\tilde\Delta} \sim 
2{\overline{\big[\frac{{\delta a}}{a}\big]}} {\bar e}
\end{equation}
we obtain
\begin{equation} {\tilde\Delta} \sim 
{\overline{\big[\frac{{\delta a}}{a}\big]}}\frac{27\hbar^2 N^2}{8m^*\pi^2} 
\end{equation}
and hence ${\tilde\Delta} \sim N^2$ as claimed.

Second, we investigate the general $N$-dependence of $E_0$. Following Eqs. (47)
and (48), we assume that $E_0$ is dominated by the classical potential energy at
the SEM $\Omega'\equiv\Omega'_0$, i.e. $E_0\sim \beta V(\Omega'_0)$. This is
consistent with our assumption throughout the paper of considering configurations
close to the classical minima.  Hence  \begin{equation}
E_0\sim\beta\sum_{j<j'}\frac{1}{|\alpha_{j0}e^{i\theta_{j0}}-
\alpha_{j'0}e^{i\theta_{j'0}}|^2}\ . \end{equation}    Replacing the denominator
by $\overline{{\alpha}^2}$ yields
$E_0\sim\beta\frac{1}{\overline{{\alpha^2}}}\sum_{j<j'} 1$ and hence $E_0\sim \beta
N\frac{1}{2}(N-1)(N-2)$. For large $N$, $E_0\sim \beta\frac{N^3}{2}$ and hence
$E_0\sim N^3$ as claimed. We note that while this derivation is crude, the final
expression for $E_0$ is not actually used in the calculation of energy gaps. The
only result used is the conclusion that $E_0>>{\tilde\Delta}$ for large $N$.

 \newpage
\centerline{\bf Figure Captions}

\bigskip

\noindent Figure 1:  Jacobi coordinates for the $N=3$ electron system.  Reading
clockwise, the classical configuration for three  repulsive particles  
$\Omega'_0\equiv (132)$ corresponds to 
$(\alpha,\theta)=(\frac{\pi}{4},\frac{\pi}{2})$  (i.e. $(x,y)=(0,0)$); the other
classical configuration 
$\Omega'_1\equiv (123)$
corresponds to  $(\alpha,\theta)=(\frac{\pi}{4},-\frac{\pi}{2})$ (i.e.
$(x,y)=(0,\pi)$ or, equivalently, $(0,-\pi)$). 

\bigskip
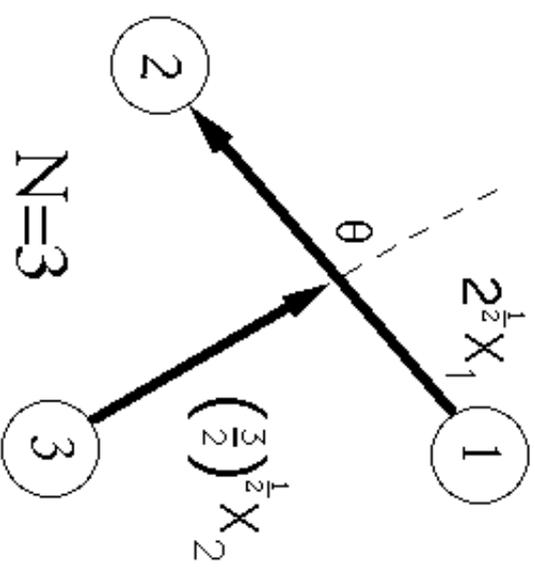
\noindent Figure 2: Schematic diagram showing a portion of the hyperangular
$\Omega'$-space. Two symmetrically equivalent minima (SEMs -- labelled as
$\Omega'_i$) are indicated by larger circles, while nearby symmetrically
inequivalent minima (SIMs -- labelled as $\Omega'_{i;a}$) are indicated by smaller
circles.

\bigskip
\noindent Figure 3:  Contour plot of ficticious potential $ V(x,y;\epsilon)$  in
the $(x,y)$ plane  for the $N=3$ electron problem. The two symmetrically equivalent
minima (SEMs -- $\Omega'_0$ and $\Omega'_1$ from Fig. 1) are shown. Minima in
$V(x,y;\epsilon)$ occur at $(0,0)$ and $(0,\pm\pi)$  (i.e. at the classical 
configurations).  Maxima occur at $({\rm ln} {\sqrt 3}, \pm\frac{\pi}{2})$,  where
$V(x,y;\epsilon)\rightarrow\infty$ (i.e. electrons 2 and  3 or 1 and 3
coincident).  $V(x,y;\epsilon)$ is positive and finite everywhere else.  The same
qualitative features appear for all $\epsilon$  ($\frac{\epsilon}{m^*\beta}=5$ is
used as an illustration). 

\bigskip
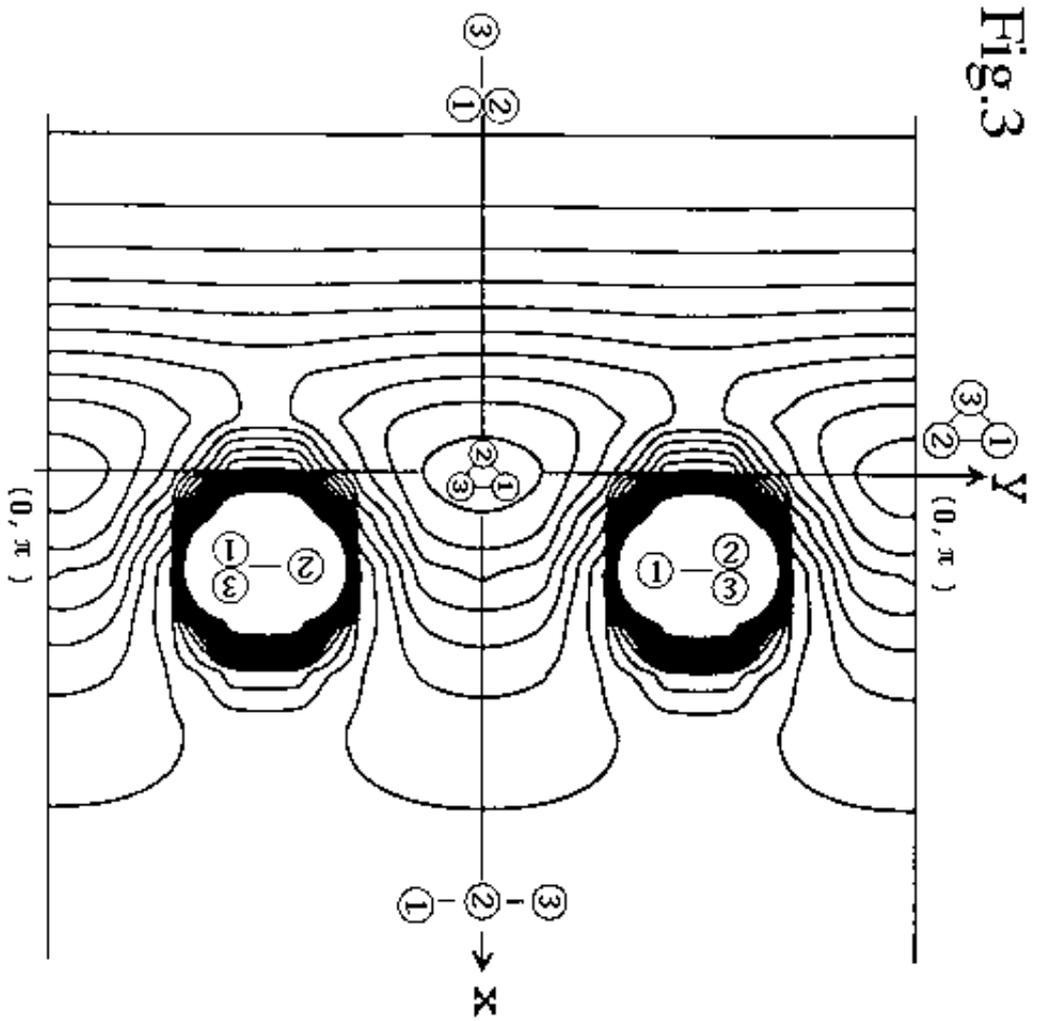
\noindent Figure 4: Ground-state classical configuration for 230 electrons (black
dots) calculated numerically using a Monte Carlo simulation. Figure adapted from
Fig. 2 of Bedanov and Peeters (Ref. 14). Straight lines
are drawn to bisect the midpoint between nearest neighbor electrons, thereby
highlighting the approximately hexagonal local symmetry.  Circles are drawn to
illustrate  the approximately ring-like arrangement of electrons. The inner circles
only pass through regular hexagons. The outer circle passes through several
pentagons and distorted hexagons because of its proximity to the edge of the finite
cluster. 

\bigskip
\noindent Figure 5: A particular symmetrically equivalent minimum (SEM -- labelled
in the text as $\Omega'_0$) with
$N$ near the center (c.f. Fig. 4). We are considering the limit of large $N$.
Ring
$m$ is such that
$N>>j>>1$ and it is therefore far away from the circumference of the
cluster.  

\bigskip
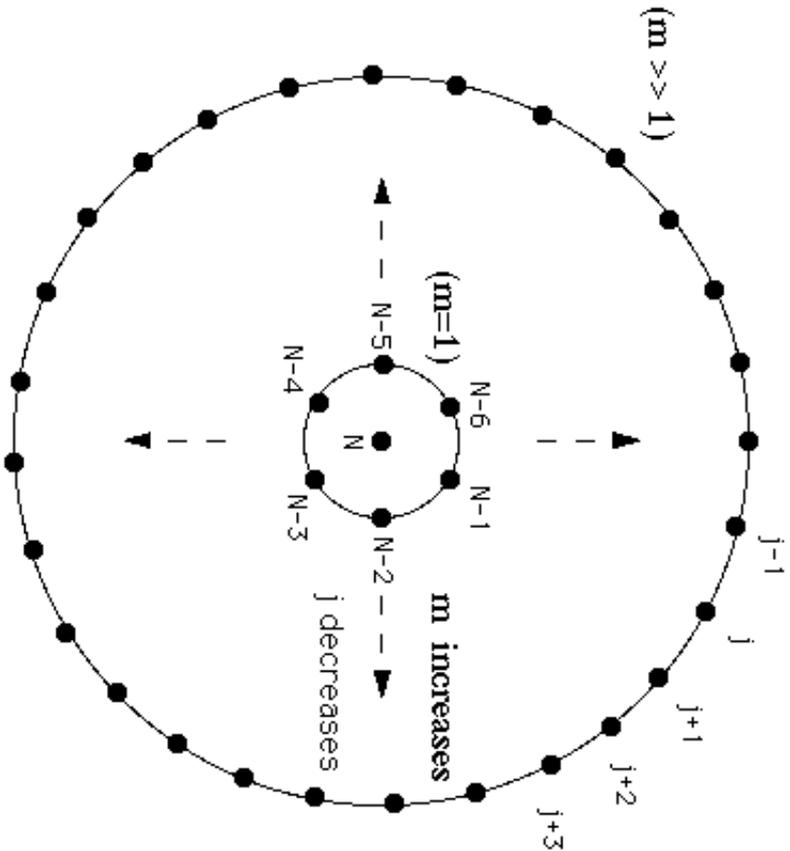
\noindent Figure 6: Typical ring $m$ for the various spin-polarizations. Although
only $12$ particles are shown for clarity, the ring is assumed to contain a large
number, i.e. $N_m>>1$ since $m>>1$. (a)
$N_+:N_-=1:1$. (b) $N_+:N_-=2:1$. (c) $N_+:N_-=3:1$. 

\bigskip
\noindent Figure 7: Theoretical lower estimate (straight line) for FQHE
energy gaps obtained from the present theory as compared to experimental results
over a range of lower mobility GaAs heterostructure samples. (Figure adapted from
the data of Boebinger et al., Refs. 27 and 3). Experimental data:
$\nu=\frac{1}{3}$ (black points),
$\nu=\frac{2}{3}$ (white points). As discussed in the text, 
theoretical gaps are same for $\nu=\frac{1}{3}$ and $\frac{2}{3}$ to
a first approximation.

\bigskip
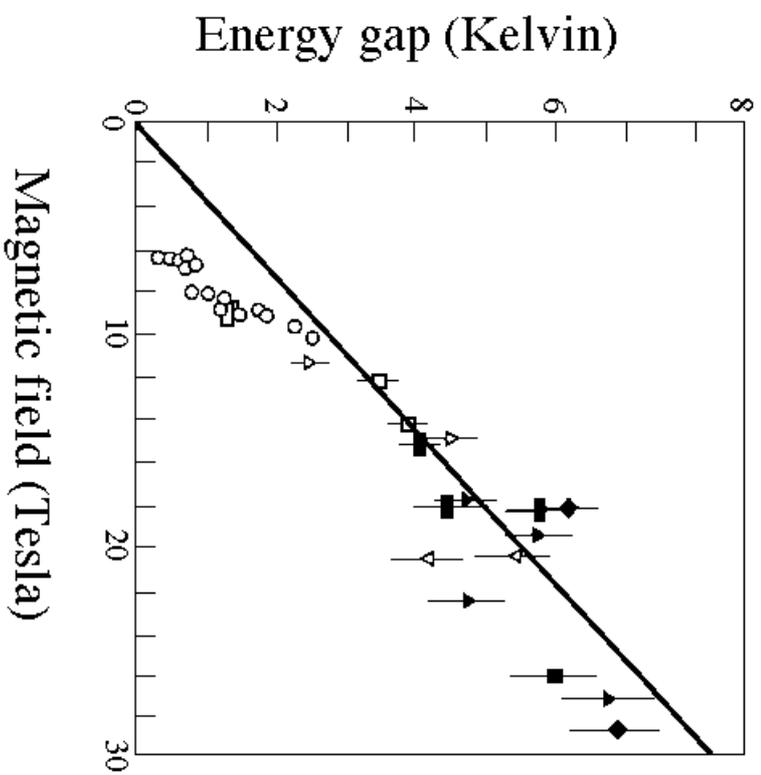
\noindent Figure 8: Theoretical upper (dashed line) and lower (solid line -- same
as Fig. 7) estimates for FQHE energy gaps obtained from the
present theory as compared to experimental results over a range of higher mobility
GaAs heterostructure samples. Data taken from Ref. 28 of Mallett et al. (squares)
and Ref. 30 of Willett et al. (solid circles
with error bars). Experimental data contains values for fractions
$\nu=\frac{p}{3}$ where
$p=1,2,4,5$. As discussed in the text, theoretical gaps are independent of $p$ to
a first approximation.


\newpage


\begin{thebibliography}{99} 
\bibitem{FQHE} R.E. Prange and S.M. Girvin, {\em The
Quantum Hall Effect}, (Springer, New York, 1990).
\bibitem{alan}A.H.
MacDonald, {\em Quantum Hall Effect: a perspective}, (Academic, New York, 1990).
\bibitem{Chakraborty} T. Chakraborty and P. Pietilainen, {\em The Quantum Hall
Effects}, (Springer-Verlag, Heidelberg, 1995).  
\bibitem{Laughlin} R.B. Laughlin, Phys. Rev. Lett.
{\bf 50}, 1395 (1983).
\bibitem{bert} B.I. Halperin, Helv. Phys. Acta {\bf 56}, 75 (1983); Sci.
Am. {\bf 254}, 52 (1986).
\bibitem{jain} J.K. Jain, Adv. Phys. {\bf 41}, 105 (1993).
\bibitem{HLR} B.I. Halperin, P.A. Lee and N. Read, Phys. Rev. B {\bf 47}, 7312
(1993).
\bibitem{ashoori} 
R.C. Ashoori, H.L. Stormer, J.S. Weiner, L.N. Pfeiffer, K.W.
Baldwin and K.W. West, Phys. Rev. Lett. {\bf 71}, 613 (1993).
\bibitem{mceuen} 
P.L. McEuen, E.B. Foxman, J. Kinaret, U. Meirav, M.A. Kastner, N.S.
Wingreen and S.J. Wind, Phys. Rev. B {\bf 45}, 11 419 (1992).
\bibitem{Maksym} P.A. Maksym and T. Chakraborty, Phys. Rev. Lett. {\bf 65},  108
(1990);  P. Hawrylak, Phys. Rev. Lett. {\bf 71}, 3347 (1993);
S.-R. Eric Yang, A.H. MacDonald and M.D. Johnson,
Phys. Rev. Lett. {\bf 71}, 3194 (1993); P. Hawrylak and D. Pfannkuche, 
Phys. Rev. Lett. {\bf 70},  485 (1993).  
\bibitem{hansen} See for example 
W. Hansen, T.P. Smith, K.Y. Lee, J.M. Hong and C.M. Knoedler, Appl.
Phys. Lett. {\bf 56}, 168 (1990).
\bibitem{ourPRL} N.F. Johnson and L. Quiroga, Phys. Rev. Lett. {\bf 74}, 4277
(1995).
\bibitem{bolton} F. Bolton and U.
Rossler, Superlatt. and Microstruct.  {\bf 13}, 140 (1993). 
\bibitem{Bedanov} V.M. Bedanov and F.M. Peeters, Phys. Rev. B {\bf 49}, 2667
(1994).
\bibitem{Eckardt} P.A. Maksym, Phys. Rev. B {\bf 53}, 10871 (1996). Some related
ideas were presented by W. Hausler and B. Kramer, Phys. Rev. B {\bf 47},
16353 (1993). See also P.A. Maksym in Proceedings of 1996 High Magnetic Field
Conference, Wurzburg (World Scientific, 1997). 
\bibitem{Kivelson} S. Kivelson, C. Kallin, D.P. Arovas and J.R. Schrieffer,
Phys. Rev. Lett. {\bf 56}, 873 (1986); Phys. Rev. B {\bf 36}, 1620 (1987). 
\bibitem{review} N.F. Johnson, J.
Phys.: Condens. Matt. {\bf 7}, 965 (1995); 
L. Quiroga, D. Ardila and N.F. Johnson, Solid State Comm. {\bf
86}, 775 (1993). 
\bibitem{Maksym96} L.D. Hallam, J. Weis and P.A. Maksym, 
Phys. Rev. B {\bf 53}, 1452
(1996).
\bibitem{madhav} A.V. Madhav and T.
Chakraborty, Phys. Rev. B {\bf 49},  8163 (1994). 
\bibitem{kinaret} 
J.M. Kinaret, Y. Meir, N.S. Wingreen, P. Lee and
X-G. Wen, Phys. Rev. B {\bf 46}, 4681 (1992).
\bibitem{pakming}
P.M. Hui (unpublished). 
\bibitem{term}
This terminology follows that introduced in molecular physics and later adopted by
Maksym in Ref. 15. See E. Bright Wilson, J. Chem. Phys. {\bf 3}, 276 (1935). 
\bibitem{fock} V.
Fock, Z. Physik {\bf 47}, 446 (1926). 
\bibitem{Geller}
M.R. Geller and G. Vignale, Phys. Rev. B {\bf 53}, 6979 (1996).
\bibitem{Schweigert}
V.A. Schweigert and F.M. Peeters, Phys. Rev. B {\bf 51}, 7700 (1995).
\bibitem{Fisher} D.S. Fisher, B.I. Halperin and R. Morf, Phys. Rev. B {\bf 20},
4692 (1979).
\bibitem{Boebinger} G.S. Boebinger, H.L. Stormer, D.C. Tsui, A.M. Chang, J.C.M.
Hwang, A.Y. Cho, C.W. Tu and G. Weimann, Phys. Rev. B {\bf 36}, 7919 (1987).
\bibitem{Robin} J.R. Mallett, R.G. Clark, R.J. Nicholas, R. Willett, J.J. Harris
and C.T. Foxon, Phys. Rev. B {\bf 38}, 2200 (1988). 
\bibitem{Robin2} D.R. Leadley, M. van
der Burgt, R.J. Nicholas, C.T. Foxon and J.J. Harris, Phys. Rev. B {\bf 53}, 2057
(1996). 
\bibitem{Willett} R.L. Willett, H.L. Stormer, D.C. Tsui, A.C. Gossard and J.H.
English, Phys. Rev. B {\bf 37}, 8476 (1988). 
\end{thebibliography}
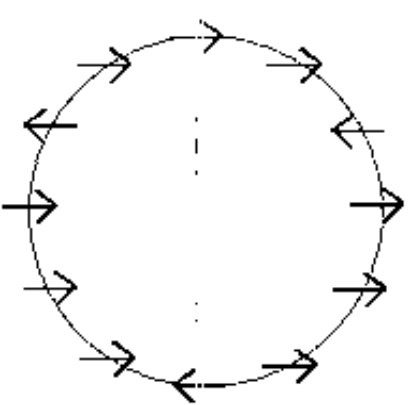
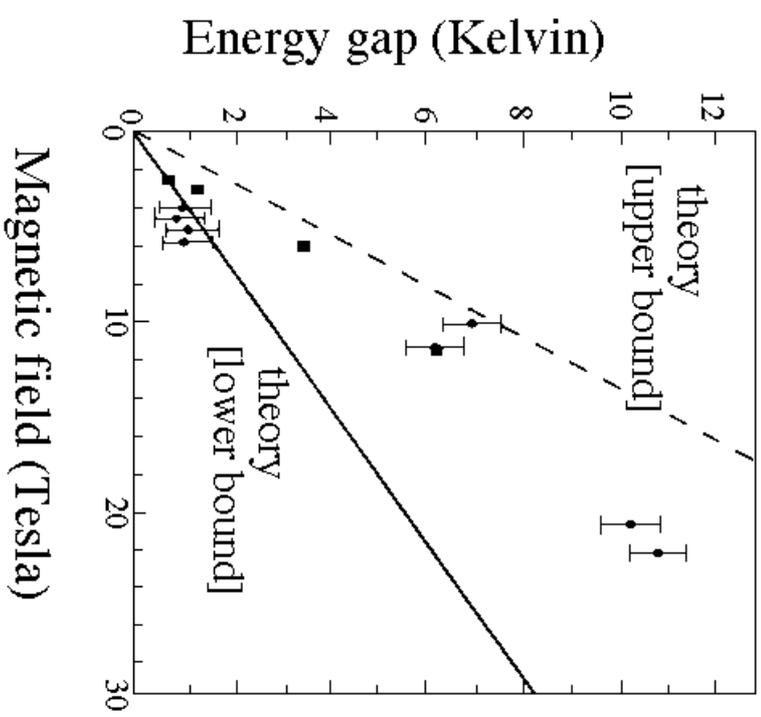
\end{document}